\documentclass[12pt]{article}
\usepackage[utf8]{inputenc}

\usepackage{graphicx}
\usepackage{natbib}
\usepackage{fullpage}
\usepackage{rotating}                                       
\usepackage{amsmath}
\usepackage[table,xcdraw]{xcolor}                           
\counterwithin*{equation}{subsection}                       

\usepackage[section]{placeins}                               
\usepackage{amssymb}
\usepackage{subcaption}
\usepackage[labelfont=bf]{caption}
\usepackage[linewidth=1pt]{mdframed}

\graphicspath{{fig/}}
\title{\vspace{-1.5cm}\noindent\makebox[\linewidth]{\rule{\textwidth}{0.4pt}}\vspace{8mm} \Huge{\textbf{Reconstructing missing seismic data using Deep Learning}}}
\date{}
\author{
	\Large{Dieuwertje Kuijpers\textsuperscript{1}, Ivan Vasconcelos\textsuperscript{1} and Patrick Putzky\textsuperscript{2}} 
    \vspace{5mm} \\
	\textsuperscript{1}{Utrecht University, Utrecht, the Netherlands}\\ 
	\textsuperscript{2}{AMLab University of Amsterdam, Amsterdam, the Netherlands}
}

\usepackage{natbib}
\usepackage{graphicx}

\begin{document} 
\maketitle
\thispagestyle{empty}
\noindent\makebox[\linewidth]{\rule{\textwidth}{0.4pt}}

\section*{\centering Abstract}
\hspace{\parindent} In current seismic acquisition practice, there is an increasing drive for data to be acquired sparsely in space, and often in irregular geometry. These surveys can trade off subsurface information for efficiency/cost - creating a problem of ``missing seismic data'' that can greatly hinder subsequent seismic processing and interpretation. Reconstruction of regularly-sampled dense data from highly-sparse, irregular data can therefore aid in processing and interpretation of these far sparser, more efficient seismic surveys. Here, we compare two methods to solve the reconstruction problem in both space-time and wavenumber-frequency domain. Both of these methods require an operator that maps sparse to dense data: this operator is generally unknown, being the inverse of a known data sampling operator. As such, here our deterministic inversion is efficiently solved by least squares optimisation using an numerically-efficient Python-based linear operator representation. An alternative method is the probabilistic approach that uses deep learning. Here, two specific deep learning architectures are benchmarked against each other and the deterministic approach; a Recurrent Inference Machine (RIM), which is designed specifically to solve inverse problems given known forward operators and the U-Net, originally designed for image segmentation tasks. The trained deep learning networks are capable of successfully mapping sparse to dense seismic data for a range of different datasets and decimation percentages, thereby significantly reducing spatial aliasing in the wavenumber-frequency domain. The deterministic inversion on the contrary, could not reconstruct the missing data in space-time domain and thus did not reduce the undesired spatial aliasing. Our results show that the application of Deep Learning for the seismic reconstruction is promising, but the treatment of large-volume, multi-component seismic datasets will require dedicated learning architectures not yet realisable with existing tools. 

\newpage\cleardoublepage
\setcounter{page}{1}
\section{- Introduction}
Efficient and cost-effective data acquisition is, together with streamlined data processing, of crucial importance in seismic imaging, from exploration to the global scale. In the example of exploration surveys, acquisition is designed to sample data at a set Nyquist rate (or higher), driving costs to be very high and the duration to often be very long. In principle, a more beneficial acquisition model would be to use fewer sources and/or receivers, yet still maintaining the same information content as a more conventional high-density, regularly-sampled setup. However, on its own, sparse, irregular acquisition results in missing data/information due to sparser sampling (i.e. sub-Nyquist sampling). Missing seismic data, either due to sparser sampling or irregularities can greatly hinder accurate processing and interpretation.  For example~\cite{peng2019study} find that missing seismic data in either source or receiver domain or both domains can lead to different types of artifacts and data gaps after using the sparse datasets for Marchenko methods. The reconstruction of dense, regularly sampled wavefields from highly sparse, (ir)regular data can therefore play a critical role in achieving better processing and interpretation from far sparser, more efficient seismic surveys. 

Several methods exist to solve this reconstruction problem. These methods can broadly be divided into two groups; deterministic and probabilistic. Most often the reconstruction problem is solved using deterministic, iterative linear solvers. \cite{ruan2019data} for example, find that the sampling rate in seismic acquisition can be decimated further than the Nyquist rate by means of preconditioning and compressive sensing techniques in the presence of acquired data gradients. Using a multi-component reconstruction theorem that includes the acquired data, the first- and second-order spatial derivatives plus the cross-derivatives in shot- and receiver-domain, \cite{JMthesis} can succesfully reconstruct regularly decimated 3D seismic data with one-third of the original Nyquist rate using a gradient-based, sparsity promoting solver. When using an irregular sampling scheme as proposed by \cite{hennenfent2008simply}, \cite{JMthesis} can decimate the sample rate even further. One major requirement for this method is the need for spatial derivatives of the data in the inversion: in practice, this would mean that data are acquired far more sparsely, but each data station contains many channels due to the multi-component nature of gradient data. For example, in offshore seismic, derivatives of the wavefield can be measured if particle-velocity measurements are available, something that is often not the case for vintage seismic data and also presents technological challenges in practice, such as the engineering of source-side derivatives, or higher order derivatives on either source or receiver side. 

The interest in machine learning solutions to inverse (seismic) problems is growing, the reconstruction problem provides an attractive application because the underlying forward operators are computationally inexpensive. For deterministic approaches however, achieving accurate solutions to data reconstruction can be quite challenging. Recently, \cite{siahkoohi2018seismic} addressed the use of adversarial neural networks (GANNs) to learn a map from sparsely to fully sampled seismic data. With the use of their trained GANN, \cite{siahkoohi2018seismic} are able to reconstruct 90 percent of the missing seismic data in frequency domain under different types of frequency domain decimation, as long as at least 5 percent of the data in that particular frequency slice was densely sampled. Seismic acquisition however, is often done in the spatial domain and thus does the decimation also takes place in the spatial domain. 

This research will focus on reconstructing dense seismic wavefields from spatially decimated data using deep learning, by means of the so-called Recurrent Inference Machine (RIM) deep learning architecture designed by \cite{putzky2017recurrent}. Testing the potential of using RIMs in seismic processing problems where determining a complex inverse map to a known forward problem is the main goal. The RIM will be benchmarked against the U-Net deep learning architecture (originally designed for biomedical image segmentation; \cite{ronneberger2015u}) and will be compared to deterministic linear iterative methods. 

Deep learning mainly consists of two stages. The first stage is the training stage in which the neural networks have access to an input and expected output. Based on the input the network has to make a prediction that should be as close as possible to the expected output. The misfit between the prediction and expected output can be backpropagated through the network thereby updating its internal state in order to make a better prediction for the next example. After a period of training, the neural nets enter the inference stage. In this stage the network will have access to input data, that it has never seen before, only. From this input the network should try to make a prediction. Here, the reconstruction problem will be studied and the neural networks will estimate a map between the decimated and dense seismic wavefields in which deep learning can be seen as an approach to solving inverse problem. 

The reconstruction problem will be studied in the time-space domain mostly as most seismic data are acquired in this domain. In the frequency-wavenumber domain the reconstruction problem becomes the dealiasing problem as sub-Nyquist spatial sampling will lead to spatial aliasing. After studying the approach the two methods take in solving inverse problems, the reconstruction problem will first be studied in 2D where decimation (with different patterns and percentages) only takes place along the receiver dimension. As a final test all different studied methods will aim at solving a highly decimated 3D Ocean Turbulence dataset, that is not just decimated along the receiver dimension but also along the source dimension, resulting in over 90 \% missing data to be reconstructed. The next section gives the reader a general introduction to machine learning, a deeper description of the specific architectures used here will be given in coming sections. 

\newpage\cleardoublepage
\section{- A brief introduction to Machine Learning}

In this section, a short introduction to machine learning is given to help the reader understand the techniques used in this research. Because the machine learning community often uses a specific wording that will also be used in this study, a short glossary is given at the end of this section. The introduction and glossary are far from complete as they only serve to describe the basic concepts. Two recommended references for a more detailed description or a more hands-on experience include the book on Deep Learning by \cite{Goodfellow-et-al-2016} and the online course on Pytorch via \cite{udacitycourse}.

A machine learning algorithm is able to learn from example data, with learning being described as an increased performance over repetitive execution of a given task. In its very mathematical basics, machine learning can be seen as a form of applied statistics since computer models are used to statistically estimate a unknown, often complicated function that maps a given input to a given output. Deep learning is a form of machine learning in which a deep (multiple layers) neural network is the learning computer model. The network is a numerical representation of a series of computations that process information. With every pass through a layer mathematical computations are applied to the input data, thereby mapping part of the input data to a new representation. The visible input and output to a machine learning network can have very different forms such as images, text or classification labels. All layers in between hold hidden representations of the data that are invisible for the user. 

The layers in a neural network consist of nodes, each different node applies a mathematical function to part of the input data. The output of each node has a different importance in the layer's representation of the data and therefore all nodes have a corresponding weight. When building a machine learning model, the weights have an initial setup that is not optimal in mapping the input to output. Thus, for a model that should generalize well to different and highly variable data, it is important to find the optimum set of weights (high weights corresponding to more import features) that represent a map between the data in a so-called training dataset. 

The network, mathematically represented by $g$, defines a parametric model between the output $\mathbf{\tilde{x}}$ and input $\mathbf{y}$ as set by the weights such that $\mathbf{\tilde{x}} \ = \ g(\mathbf{y}, \mathbf{w})$.  Training consists of estimating the network weights $\mathbf{w}$ by minimization of a specific loss function suitable for the problem. Training data consists of a large set of data for which both $\mathbf{x}$ and $\mathbf{y}$ are known such that the difference (loss) between model output $\mathbf{\tilde{x}}$ (generated by the network from input $\mathbf{y}$; indicated by a tilde) and, during training known, $\mathbf{x}$ can be minimized. Minimization of the loss by altering the weights during training is achieved with the help of an optimizer that performs iterative optimisation using stochastic gradient descent. The training stage is followed by the inference stage during which the trained network is deployed for testing. In this phase never before seen data $\mathbf{y}$ can be used as an input and the model will map this to a new output representation $\mathbf{\tilde{x}}$. 

A deep learning model is build by selecting an architecture suited for the specific problem, a loss function and an optimizer. Many different combinations of these three exist and here we have chosen to use convolutional networks to solve a regression problem. The most simple form of a regression problem consists of finding the parameters $a$ and $b$ fitting a linear trend ($y \ = ax + b$) with (training) data in Cartesian space. In this study the problem is more complex, the convolutional networks will take corrupted (decimated) 2D seismic gathers as input and the network should map these to an output consisting of 2D reconstructed (dense) gathers. Convolutional networks (CNNs) are capable of taking N-dimensional images as input without having to transform these into 1-dimensional vectors (a very common technique in machine learning), thereby more successfully capturing the spatial and temporal dependencies in the data. In CNNs, 2D convolutional kernels are applied to the input data, therefore the weights in a CNN correspond to kernel weights that extract higher-level features from the input. 

The main goal in deep learning is thus to find a "different" (the meaning of different is unique for each problem) representation of the input data after a forward pass through the model. The mapping function that takes input to output is then represented by the network weights. The problem of mapping corrupted to reconstructed seismic gathers can be cast as an inverse problem (forward problem: $\mathbf{y} \ = \ \mathbf{A}\mathbf{x}$) where the task is to find $\mathbf{x}$ (reconstructed gather) given $\mathbf{y}$ (corrupted gather) and the forward operator $\mathbf{A}$. In this example the weights of the neural network, representing the mapping function, should represent the inverse of the forward operator that maps $\mathbf{y}$ back to $\mathbf{x}$. Therefore, deep learning will be used in this study as a probabilistic approach to inverse problems. After the machine learning glossary, the next sections will describe the exact deep learning architectures used in this study and how each of those approach inverse problems. 

\newpage
\subsection*{Machine Learning Glossary}
\begin{itemize}
\itemsep0em
\item \textbf{Activation function} - the function applied to the input data in a node activating that node or transforming input to output. Here, the Rectified Linear Unit (ReLU) activation ($\text{ReLU}(x) \ = \ \text{max}(0, x)$) is used. 
\item \textbf{Batch} - the set of data(-patches) that is used for one update step during training of the network.
\item \textbf{Channels / Features} - features are the properties or characteristic phenomenons of the input data that are extracted in a layer. Channels and features refer to the same dimension in the data (e.g. a grayscale image consists of 1 channel and a color scale image of 3 for RGB).
\item \textbf{Dropout} - layer that randomly sets some nodes to zero during the update step in training, could help prevent overfitting.
\item \textbf{Epoch} - the time the network needs to see all training data once.
\item \textbf{Gated Recurrent Unit (GRU)} - Gating mechanism in recurrent neural networks that has feedback connections and can process entire data sequences at once. The cell regulates information flow through the network with the use of a forget and memory gate. 
\item \textbf{Learning rate} - parameter that controls the step size in stochastic gradient descent; how much the weights are adjusted with respect to the loss gradient. 
\item \textbf{Loss} - cost function that measures the misfit between the networks predictions and the expected results, loss should be minimized during the training phase.
\item \textbf{Optimizer} - the algorithm that is used to update the weights and/or learning rate in order to reduce the loss during the training phase. 
\item \textbf{Overfitting} - when an algorithms is overfitting the training data, the model remembers the output with the input instead of learning. The model therefore generalizes poorly to unseen datasets during the inference stage. 
\item \textbf{Training / Inference} - the training phase is the phase in which a machine learning algorithm is build, inference uses this trained model to make a prediction. 
\end{itemize}

\newpage\cleardoublepage
\section{- The Reconstruction problem \label{sec:recproblem}}
In sparse seismic wavefield acquisition, the reconstruction problem can be posed as a general linear problem \eqref{eq:lp};

\begin{equation}
    \mathbf{y} \ = \ \mathbf{R} \ \mathbf{x} \label{eq:lp}
\end{equation}

\noindent in which $\mathbf{y}$ is the decimated (corrupted) wavefield and $\mathbf{x}$ the dense wavefield. $\mathbf{R}$ is the Restriction operator that can be assembled from the characteristics of the acquisition setup (e.g. malfunctioning receivers or missing shots). $\mathbf{R}$ represents a mask that extracts a subset of data from the dense wavefield into the decimated wavefield. Equation \eqref{eq:lp} is known as the forward problem that generates the observed data. The inverse problem consists of reconstructing the dense wavefield $\mathbf{x}$ from the observed decimated wavefield $\mathbf{y}$ using an inverse of the restriction operator. 

From Nyquist-Shannon's sampling theorem it is known that the restriction operator in equation \eqref{eq:lp} has an exact inverse as long as the sample-rate criterion is satisfied. A main assumption in Nyquist-Shannon's sampling theorem is that of uniform sampling. In reality however, irregularities in the acquired data could be caused by malfunctioning receivers or perturbations leading to a varying receiver spacing or sample rate during acquisition. Irregular and/or far sparser sampling both result in ill-posedness of the inverse of equation \eqref{eq:lp}. In these cases the inverse of the restriction operator can be approximated by two types of approaches; iterative deterministic or probabilistic inversion. In what follows, each densely sampled gather is represented by $\mathbf{x}$ and the decimated version by $\mathbf{y}$. The goal is to estimate a dense version of the data from the decimated data and the forward operator, this estimate is represented by $\mathbf{\tilde{x}}$ and should be as close to the original dense data $\mathbf{x}$ as possible. The seismic data could be decimated over a single source- or receiver-dimension resulting in the reconstruction of missing traces in 2D seismic gathers, or decimated in both dimensions resulting in a highly sparse 3D decimated dataset. 

\subsection*{Deterministic - Linear Solvers}
Deterministic methods aim at inverting equation \eqref{eq:lp} without explicitly using any probability theory on the parameters of the inversion. The most general solution to this inverse problem is the least-squares solution to which possible regularization terms can be added. Minimizing the least squares cost function, yields the reconstructed dense wavefield $\mathbf{\tilde{x}}$ of equation \eqref{eq:DetInv}. The linear system in equation \eqref{eq:lp} can numerically be represented using an efficient linear operator representation in the Python-based Pylops framework \citep{ravasi2020pylops}. Pylops-implemented least squares optimisation can also be used to efficiently solve the inversion in equation \eqref{eq:DetInv}. Least squares optimisation uses the forward operators in the inversion and is therefore controlled by the physics of the restriction operator. 

\begin{equation}
\tilde{\mathbf{x}} \ = \min_{\mathbf{x}} || \mathbf{y} \ - \ \mathbf{R \ x} || ^2 = \ (\mathbf{R}^T \mathbf{R})^{-1} \mathbf{R}^T \mathbf{y} \label{eq:DetInv}   
\end{equation}


\subsection*{Probabilistic - Deep Learning}
An alternative method to solve the inverse problem makes use of deep learning. The neural network (mathematically represented by $g_{\phi}$) is trained to represent an approximate inverse of the restriction operator thereby mapping the decimated to the dense data. From now on $\phi$ will be used to represent the network's parameters instead of the earlier introduced $\mathbf{w}$. This because $\phi$ includes the weights and can also, since the used models are more complex than simple linear regression, include other trainable parameters like a varying learning rate. The neural network is trained to minimize the mean squared cost function $\mathbf{J}$ (see equation \eqref{eq:dlCost}) with the use of an optimizer that performs gradient descent on this cost function and the model parameters. The main focus of this study lies on the Recurrent Inference Machine (RIM) as designed by \cite{putzky2017recurrent}, which will be benchmarked to a more simplistic network architecture; the U-Net as first designed by \cite{ronneberger2015u}. The numerical code used for U-Net is based on that of \cite{zbontar2018fastmri} for their fastMRI challenge. Both existing code basements for the RIM and U-Net will be adjusted for the specific goal of reconstructing missing seismic data. 

\begin{equation}
    \mathbf{J} \ = \ || \mathbf{x} - \mathbf{\tilde{x}}||^2 \ = \ || \mathbf{x} - g_{\phi}(\mathbf{y})||^2 \label{eq:dlCost}
\end{equation}

\subsubsection*{Probability Theorem}
The parameters in a neural network should represent an unknown map between an input $\mathbf{y}$ and an output $\mathbf{x}$, that is supposed to be an inverse to a known forward operator (linear or non-linear) mapping $\mathbf{x}$ to $\mathbf{y}$. This means that the goal to solving inverse problems using deep learning comes down to creating a function estimator of the actual inverse operator. The neural network parameters are trained to represent this function estimator, the belief that these parameters ($\theta$) can represent the inverse operator can be expressed using probabilities. Maximum probability corresponds to a 100 \% capability of the network parameters to represent the desired inverse operators. Different approaches can be taken to maximize this probability (refer to Chapter 5 of \cite{Goodfellow-et-al-2016}). Here, the inverse problem is approached by defining a likelihood and a prior and optimizing the maximum a posteriori solution (MAP) in the following equation,

\begin{equation}
    \tilde{\mathbf{x}} \ = \ \max_{\mathbf{x}} \ \log p(\mathbf{y} | \mathbf{x}; \theta) \ + \ \log p_{\theta} (\mathbf{x}) \, .\label{eq:MAP} 
\end{equation}

\noindent such that the iterative approach to MAP inference represents the iterative approach to inversion (an optimization problem). 

In equation \eqref{eq:MAP}, the first term is a conditional probability (log-likelihood term) under network parameters $\theta$ that represents the forward problem, while the latter is a parametric prior over $\mathbf{x}$ that reduces the ill-posedness of the inverse problem by including for example a sparsity promoting term \citep{putzky2017recurrent}. Maximizing the conditional log-likelihood term is an attempt to make the network parameters match the mapping function between input and output as set by the training data. Ideally this would match all data used during inference, however these data are not directly available and therefore that probability distribution remains unknown. The conditional log-likelihood term is the basis for supervised learning in which $\mathbf{y}$ is predicted given $\mathbf{x}$ and the model parameters. The maximum a posteriori approach also includes the prior on the dense wavefield thereby allowing the network parameters (and therefore the estimate of the inverse function) to be affected by prior beliefs. The prior distribution is also related to the training data. In the case of seismic data, the prior space can include information on spatial and temporal signal distribution, curvature and sparsity. The next sections will describe two specific architectures used in this study and how each of those approximate the inverse problem. 

\newpage \cleardoublepage
\section{- The Recurrent Inference Machine}
By design, a Recurrent Inference Machine \citep{putzky2017recurrent}, or RIM, uses a recurrent neural network (RNN) as a recurrent approach to MAP inference. \cite{putzky2017recurrent} stepped away from taking the usual deep learning approach in which the prior and log-likelihood are learned separately and instead setup a RNN that jointly learns inference and a prior. The RIM uses the current reconstruction ($\mathbf{\tilde{x}}_t$), a hidden memory state ($\mathbf{s}$) and the gradient of the log-likelihood term ($\mathbf{\nabla} \log p(\mathbf{y} | \mathbf{x}; \theta)$) to infer a better reconstruction ($\mathbf{\tilde{x}}_{t+1}$) over a fixed number of steps in the recurrent part of the RIM. Each consecutive estimate of the recurrent part in the RIM $\mathbf{x}$ can, in its most simple form, be estimated through a recursive update function 

\begin{equation}
    \mathbf{\tilde{x}}_{t+1} \ = \ \mathbf{\tilde{x}}_t \ + \gamma_t \nabla \big( \log p(\mathbf{y} | \mathbf{x}) + \log p_{\theta} (\mathbf{x}) \big)  \, . \label{eq:updateRIM1}
\end{equation}

\noindent Using Bayes' rule and generalization to the RIMs formulation this results in recursive update equation \eqref{eq:updateRIM}. The learnable parameters $\phi$ in the RIM (represented by $g_{\phi}$ in \eqref{eq:dlCost}) now include network and prior parameters $\theta$ and the learning rate. For a more detailed description on RIMs and the derivation from equation \eqref{eq:updateRIM1} to \eqref{eq:updateRIM}, the reader is referred to \cite{putzky2017recurrent}. For now it suffices to know that the inputs to a RIM consist of a memory state, the gradient of the likelihood term (as given by the forward operator $R$) and the current reconstruction. The gradient of the likelihood term for general inverse problems where $\mathbf{y} \ = \ \mathbf{Ax}$ can be written as $\log p(\mathbf{y} | \mathbf{x}) \ = \ \mathbf{A}^T (\mathbf{y} - \mathbf{Ax})$. Because the forward operator $\mathbf{R}$ is self-adjoint, the gradient can here be written as $\log p(\mathbf{y} | \mathbf{x}) \ = \ \mathbf{R} \mathbf{x} - \mathbf{y}$.

\begin{equation}
    \mathbf{x}_{t+1}^{RIM} \ = \ \ \mathbf{x}_t^{RIM} + \ g_{\phi} \ \big(\nabla \log p(\mathbf{y} | \mathbf{x}) ( \mathbf{x}_t^{RIM}) \ , \  \mathbf{x}_t^{RIM} \ , \  \mathbf{s}_{t+1} \big) \, . \label{eq:updateRIM}
\end{equation}

\subsection*{RIM architecture}
\begin{figure}[h]
    \centering
    \includegraphics[scale = 0.45]{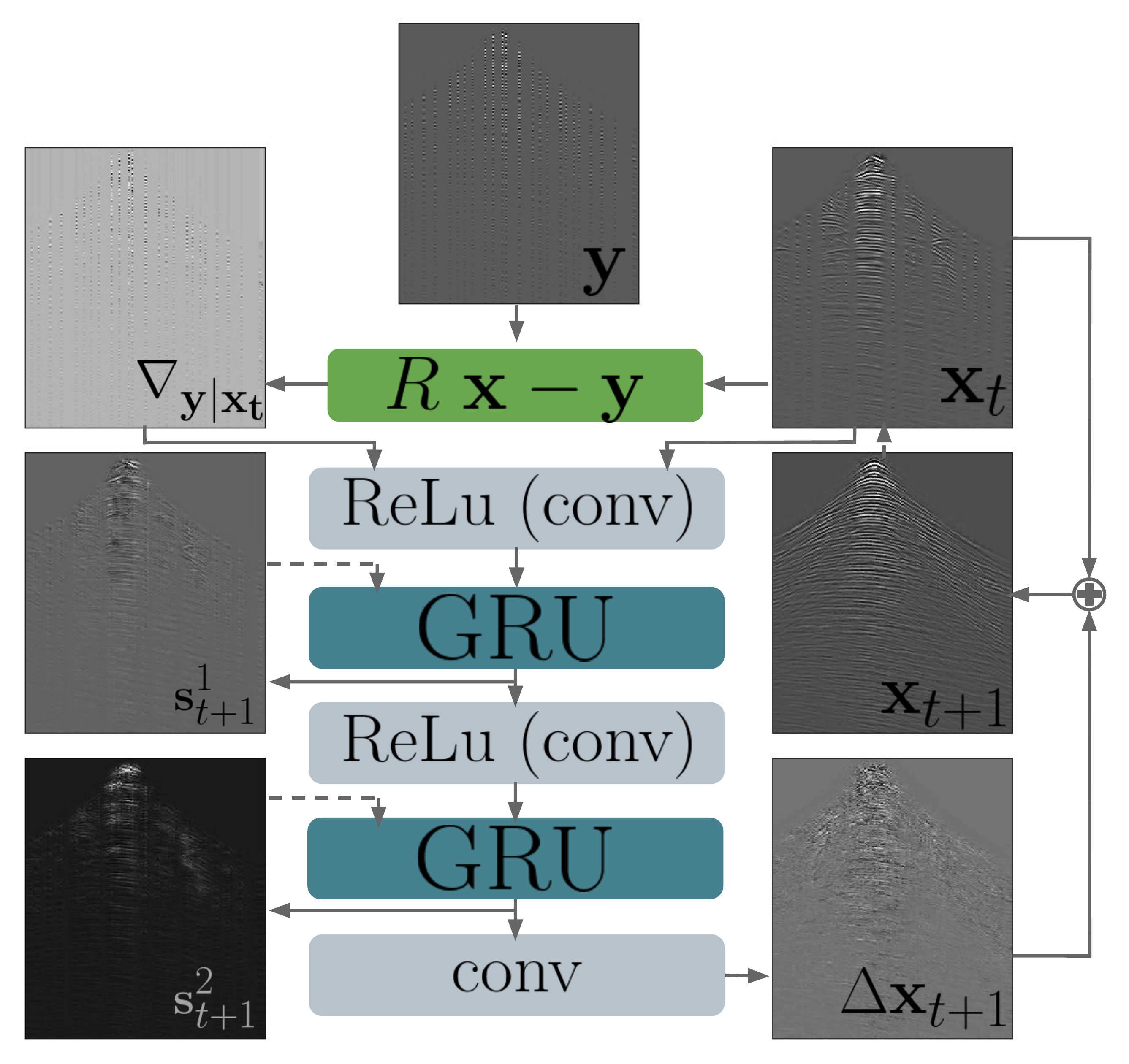}
    \caption{\textit{RIM Architecture -} An overview of the data flow through the RIM used in this project. Bold arrows are direct connections within a single timestep, dotted lines are recurrent connections passing information through to the next time step. \textit{Conv} is short for convolution and $\mathbf{\nabla_{y|x_t}}$ for the gradient of the log-likelihood term. The different representations of the input data throughout the model are described in the main text. \textit{Figure adapted from \cite{lonning2018recurrent}}.}
    \label{fig:rim}
\end{figure}

The RIM can be seen as a series of repeating neural nets configured in a single cell representing the iterative approach to inverse problems (indicated by subscripts $t$ and $t+1$ in figure \ref{fig:rim}). The RIM cell consists of a Gated Recurrent Unit (GRU) and convolutional layers. The flow through a cell is intrinsically repeated by a fixed number of steps (here chosen to be 10). Over these steps the network should improve its reconstruction for which it uses an intrinsic loss function that compares the inference prediction with the expected outcome (known for all training data). For both the intrinsic and global loss in the RIM the mean squared error is used (see equation \eqref{eq:dlCost}).

In figure \ref{fig:rim} input image $\mathbf{y}$ is the decimated data. The forward operator generating this decimated data is applied to the current estimate of the RIM ($\mathbf{x}_t$) to generate the gradient of the log-likelihood term in the green cell. The gradient (indicated by $\mathbf{\nabla_{y | x_t}}$; short for $\mathbf{\nabla} \log p (\mathbf{y} | \mathbf{x})$) and the current estimate ($\mathbf{x}_t$) of the dense wavefield, are concatenated over the channel dimension and form the input to the first convolutional layer that is followed by a ReLu activation layer. The next layer is a GRU (gating mechanism) that determines what information in the hidden state ($\mathbf{s}_{t+1}^1$) is important and what can be forgotten for the next step. Another convolutional layer followed by ReLU activation and a GRU pass (with hidden state $\mathbf{s}_{t+1}^2$) follows before the final convolutional layer. The exact RIM architecture chosen here consists of three hidden convolutional layers, the first with kernel size 5x5 and the last two having size 3x3. Padded convolution is used to have a constant image size throughout the whole network. The output in the recurrent network is an update $\Delta \mathbf{x}_{t+1}$ that is added to the current estimate ($\mathbf{x}_{t}$) to form the new estimate ($\mathbf{x}_{t+1}$). Neural networks extract features from the input to learn about data characteristics, in the first two hidden layers 64 features are extracted from the input that consists of two channels (the decimated data concatenated with the gradient of the log-likelihood term), the final output consists of a single channel; the grayscale reconstructed seismic gather $\mathbf{x}_{t+1}$, that becomes $\mathbf{x}_{t}$ in the next timestep. In total the RIM consists of just over 90.000 trainable parameters.
\newpage \cleardoublepage
\section{- U-Net}
The U-Net is a very well-known deep learning architecture for image tasks, with the benefit of being relatively easy to implement and train. The U-Net consists of a contracting path, a bottleneck in the center and an expanding path. The two paths consist of a number of blocks in which convolutional operations are applied. The contracting path maps the input $\mathbf{y}$ to a hidden representation in the bottleneck layer, thereby compressing the input to a higher-level feature representation over the blocks. The expanding path transforms the hidden representation coming from the bottleneck layer into an estimate $\mathbf{\tilde{x}}$ of the dense data $\mathbf{x}$, thereby decreasing the number of features over the blocks while increasing the size of the data. Thus, the contracting path of the U-Net is trained such that maps the corrupted input to a compact representation of the reconstructed data and the expanding path is trained to map from this compact, hidden representation to the full reconstructed data. 

What is special about the U-Net is that the features from each contracting block are concatenated to the features from the expansion block at the same level. Concatenation ensures that the learned features in the contracting path are used to build up the image in the expansion path. In contrast to the RIM, the U-Net has no knowledge of the forward operator that created the decimated data. This means that where the RIM is forced to follow the physics set by the restriction operator, the U-Net does not and that is expected to sometimes lead to physically implausible results. Here, the same loss function and optimizer as for the RIM are used. 

\begin{figure}[h]
    \centering
    \includegraphics[width=\textwidth]{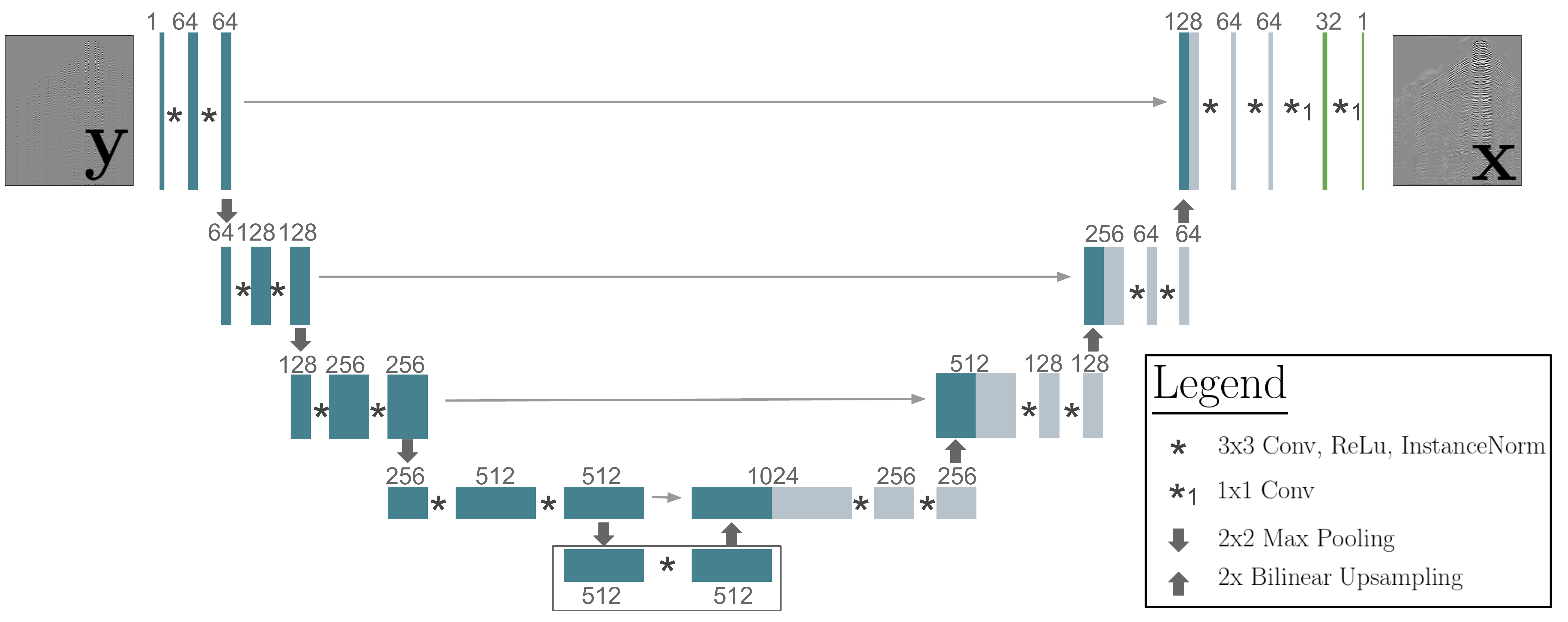}
    \caption{\textit{U-Net Architecture -} An overview of the data flow through the U-Net as used in this project, the different representations are described in the main text. The colours of the cells represent from which path the features come; blue for the contracting path, gray for the expanding path and green for the fully connected layers. \textit{Conv} is short for convolution, the numbers above the cells stand for the number of features as present in the representation of the data in that cell, width of the cell for the number of features and length for the size of the representation of the data.}
    \label{fig:unet}
\end{figure}

In the U-Net blocks, 2D max-pooling, bilinear upsampling and instance normalization are used. Pooling is a form of non-linear downsampling, the convolutional kernels output an image of the same dimensions as the input with a different number of features. Max pooling is used to reduce the size of the data between two blocks in the contracting path thereby as well reducing the required number of parameters (the more parameters the more the network is prone to overfitting the training data), memory load and number of computations. The output from one block is reassembled into small windows from which only the maximum values are kept and assembled to form the input to the next block. Pooling is a valid operation in the reasoning behind U-Net because the exact location of a feature is less important than its relative position in the global image. In order to undo this downsampling process in the contracting path, bilinear upsampling is used in the expanding path. In bilinear upsampling linear interpolation is used to interpolate the missing data in a 2D grid. First, one of the dimensions is kept fixed and linear interpolation occurs in the other direction and the second step is vice-versa. Each step is thus linear but the total interpolation is non-linear on the sampled location. Similar to the effect of data and feature normalization on network performance, instance normalization improves training by normalizing the data over the channel dimension. 

\subsection*{U-Net architecture}
The used U-Net architecture consists of four pooling blocks that perform 3x3 convolutions in both the contracting and expanding path, no dropout is used in these blocks. In figure \ref{fig:unet}, the input to the contracting path (indicated in blue) consists of a seismic gather that is decimated in the spatial domain (the same $\mathbf{y}$ as in the RIM). In the first block 64 features are extracted from the gather, this number doubles each block in the contracting path (indicated by cell width) and reaches its maximum at 1024 features in the bottleneck layer (the rectangular area in figure \ref{fig:unet}). The size of the input image decreases by a factor 2 in both image dimensions per layer (indicated by the length of the cells). Over the four expanding blocks (gray in figure \ref{fig:unet}) the number of features are decreased to 64 again and in the final two 1x1 convolutional layers (indicated in green in figure \ref{fig:unet}) this decreases to a single feature image with the same size as the original input. A 1x1 convolutional layer decreases the number of features in the representations without a change in the size. In total this U-Net consist of almost 13.5 million trainable parameters. Both the input ($\mathbf{y}$; the decimated data) and the output ($\mathbf{x}$; the reconstructed data) of the U-Net thus consist of a single feature, single channel seismic gather. The concatenation between the features from the contracting and expanding bath is indicated by the gray horizontal arrows and the combined blue/grey cells. Figure \ref{fig:unet} also justifies the name of the U-Net as the input data indeed follows a U-like flow towards the output. 

\newpage \cleardoublepage
\section{- Methods}
The inverse problem, that consists of retrieving the dense seismic wavefields from the restriction operator and the decimated data, will be solved by two approaches; deterministic inversion and deep learning. Here, the main focus lies on the RIM and the potential of the RIM to solve the reconstruction problem, as an example of an inverse problem for which the forward operator is known and computationally inexpensive. The reconstruction is benchmarked against the deterministic approach and the U-Net deep learning architecture. Eventhough the U-Net is originally designed for image segmentation \citep{ronneberger2015u}, it has lately been used for other tasks as well. For both deep learning networks many different architectures, choices of activation functions, loss functions and training data are possible. The architectures used in this study have been described in previous sections, both networks are numerically implemented using the Python-based deep learning package PyTorch \citep{paszke2019pytorch}. The most important step before deploying the neural networks in their inference stage, is training the networks on seismic data representative of the data to be inferred. The trained models can then, during the inference stage, be compared to the deterministic inversion over several tasks. The least squares optimisation in the deterministic approach is numerically implemented using the efficient linear operator representation Python-based package Pylops \citep{ravasi2020pylops}. 

\subsection*{Training networks using Seismic data}
Four different seismic datasets of different formats and sizes have been used for this study. These include the Gulf Of Suez (\textit{Gulf}) field dataset that consists of 128 shots, 128 receivers and 512 timesamples, two more complex numerical subsalt datasets (\textit{Pdat} \& \textit{Rdat}) with in total 202 shots, 201 receivers and 2001 timesamples and a 3D numerical ocean turbulence dataset (\textit{OTD}) consisting of 300 shots, 301 receivers and 1081 timesamples. A range of different networks are trained on different parts of these datasets. To generate synthetic sparser (decimated) training data for the neural networks, the originally densely sampled, in source, receiver and time domain, data are decimated using five different decimation patterns on the receiver domain. To limit the possible effects of the selected training decimation patterns on the networks capability to generalize to other decimation patterns, two jittered irregular (based on ideas of \cite{hennenfent2008simply}) and three regular (factor 2, 3 and 4) decimation patterns are applied. During training the decimation percentages vary between 50 and 80 \%. 

It is well known that sufficient data is required to accurately train a neural network (e.g. \cite{siahkoohi2018seismic}). For this study a single GPU (Nvidia GeForce GTX 1080 Ti; 11 GB memory) is used. In order to both increase the amount of training data while decreasing the computational memory load on the single GPU, non-overlapping patches consisting of 32 traces of each 64 timesamples are extracted from all shot gathers. The patches are decimated using five different masks resulting in 5 times as many decimated input wavefields as there are dense wavefields. The data-windowing effect on the data is a band-limitation of the original signal, therefore the full frequency content of the original signal is no longer present in the windowed signal. Next to that, edge effects could include undesired peaks in the frequency spectrum related to smaller-scale structures. To reduce this effect, a 2D Tukey taper (with fraction 0.3) is applied to the windowed gathers. This space-time domain multiplication of the windowed data and the Tukey taper results in a smoothing convolutional operation in frequency-wavenumber domain, that attempts to diminish the undesired effects introduced by space-time windowing. In the inference the seismic gathers will not be windowed and therefore tapering is only used in the training stage. Note thus that it is not needed to train a neural network on the same size input data as used for inference. 

\subsubsection*{Prior space sampling}
\begin{figure}
    \centering
    \includegraphics[width=\textwidth]{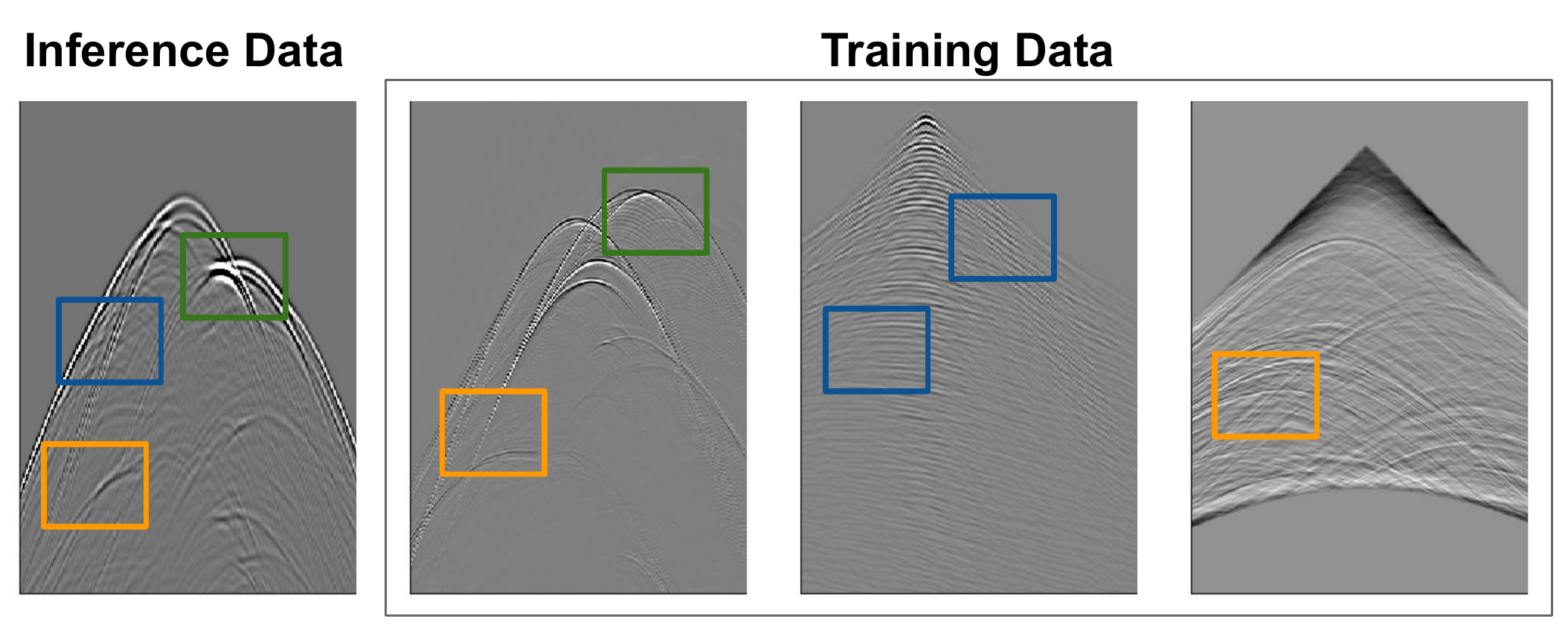}
    \caption{Illustration of the importance of a representative prior space in the training data. The to be inferred data on the left is a complex dataset consisting of many slope variations (blue), of variation in scale of structures (bandwidth; green) and a combination of diffractions and reflections (orange). The training data should therefore consist of a combination of the properties desired to be inferred by the network. All different datasets have different properties as explained in the main text. \textit{From left to right the used shot gathers are from Pdat (shot 1), Rdat (156), Gulf (47) and OTD (145).} }
    \label{fig:priorspace}
\end{figure}

To make the best predictions possible for unseen data during the inference stage, the trained deep learning algorithms require the prior space inferred from the training data to be an accurate description of the space that the networks have to infer. In the case of reconstructing seismic data, it is important for the training data to have similar slope variation, curvature and types of reflections as in the to be inferred data. Next to that, the bandwidth of the reconstruction plays an important role. The finer the temporal and spatial scale structures in the to be inferred data are, the broader the bandwidth of the training data should be. From later results it will become clear that having an idea on the decimation percentage in the data to be reconstructed can improve the network's predictions. This is related to the fact that the network's prediction quality will start to decrease at the higher end of the range of decimation percentages present in the prior. Therefore it is important to generate synthetic data with high decimation percentages for training if that is what should be reconstructed during inference. Figure \ref{fig:priorspace} illustrates this effect because if the left panel (\textit{Pdat}; single-shot salt data) were to be the goal of inference, it is important to include similar structures and properties in the training data. 

The four different datasets used in this study have different complexities. The Gulf Of Suez dataset (\textit{Gulf}) has little structural variations but includes velocity variations of the subsurface therefore having hyperbolas centered around the source location. The ocean turbulence dataset (\textit{OTD}) is the complete opposite of this because the velocities in the ocean layers have very little velocity variations but high structural variations (turbulence) therefore this dataset includes many different diffractions and reflections that can be off-centered and interfering. The \textit{Rdat} salt dataset is a synthetic dataset that includes all of the previously mentioned properties. All of these structures can be found in the single-shot \textit{Pdat} salt dataset, this data is however generated from a source within the medium and is therefore different from all other datasets that are generated by sources at the surface. 

\subsubsection*{General Deep Learning parameters}
Both networks make use of the Adam optimizer \citep{kingma2014adam} with weight decay factor 1e-8 and gradient norm 0.1. The initial learning rate is set to 1e-4 and can be altered by the optimizer. The networks are subject to the same mean squared loss function and use the Rectified Linear Unit (ReLU) activation function. During training, batches of 32 images are made over seismic shot gathers in windows of size 32x64. After the training stage, dense wavefields can be predicted for single decimated seismic gathers of varying sizes (does not have to equal the training data size). All models are trained for 40 epochs during which the loss is monitored using Tensorboard \citep{martinez2016overview}. The same decimation percentages used to decimate the training data for the RIM are used for the U-Net. 

Some machine learning architectures can be very sensitive to the scale of input data. Scaling the input data is known to have a positive effect on network performance as it is a helpful approach to the vanishing gradient problem that often occurs during back-projection of the misfit (e.g. \cite{ioffe2015batch}; \cite{dai2019channel}). The variety in amplitude and complexity of the different seismic datasets is high, scaling is therefore applied to reduce this variance and improve training. Four different types of scaling are compared; normalisation (to range -1, +1), normalisation (using maximum absolute amplitude), standardisation (zero mean, unit standard deviation) and no scaling of original data.

\subsection*{Reconstruction Approach} 
During both the training and inference stage in the deep learning approach, a single decimated 2D seismic gather is used as input. During the inference stage, the 2D decimated wavefields for unseen data map are mapped to dense reconstructions. The same synthetically generated decimated gathers, are used to perform a deterministic inversion with the help of Pylops' least squares optimisation over a 1000 iterations. The inference and inversion results will be compared over two tasks; 2D seismic gather and 3D highly decimated reconstruction. 

Unlike the deep learning networks that can only take single 2D gathers as input, the deterministic approach can invert the problem for any N-dimensional decimated data. Next to that, it is also known from compressive sensing techniques that far sparser data can be reconstructed  by inversion with the help of derivatives of the decimated data (e.g. \cite{JMthesis}). To test the potential of the neural networks (specifically trained to perform 2D reconstruction) to be used for more complex 3D highly sparse data decimated over both source and receiver domain, the 3D reconstruction problem is split into two 2D problems. First, all shot gathers will be reconstructed and after sorting the data to common receiver domain, inference can again be applied to the receiver gathers to reconstruct the rest of the missing data. This two-step approach will be compared to least squares optimisation using the first- and second-order derivative of the Ocean Turbulence data as well as the cross-derivatives in the source- and receiver-domain. The ocean turbulence dataset is a seismic dataset generated from a synthetic 2D model as described in more detail by \cite{JMthesis}. All (cross-)derivatives are created synthetically with the use of Pylops' linear operators and are decimated as well to simulate the effect of measuring these derivatives in the field with the use of streamers. 

\subsubsection*{Evaluation Metrics}
The different results will visually be compared in both the space-time (data reconstruction) and the wavenumber-frequency domain (aliasing-dealiasing problem). To quantitatively compare the different reconstruction qualities, that are scaled differently and created differently, two different metrics are used. A common evaluation metric in inversion techniques is the (normalized) root mean squared error, in image reconstruction however the structural similarity index is more common. Both metrics focus on different aspects of the reconstruction and are here used jointly to compare the performance of inversion and inference. 

The root mean squared error (RMSE) measures the difference in per-pixel amplitude between the reconstructed and reference image thereby representing the Euclidean distance between two images. The RMSE (see equation \eqref{eq:mse}) is very easy to implement as the mean squared error is already used as the loss function in the RIM and U-Net. However, RMSE lacks the ability to use overall image structure because the comparison is made per-pixel. The Structural Similarity Index (SSIM; \cite{ndajah2010ssim}) however uses the structural properties of an image and can be computed at different local patches of the image data with the use of a sliding window. SSIM is used here as defined in equation \eqref{eq:ssim}. In which the average pixel intensities ($\mu$), their variance ($\sigma^2$) and two stabilizing factors ($c$) are used to calculate the structural similarity between two seismic gathers. 

\begin{align}
    \text{RMSE} (\tilde{x}, x) \ &= \ \sqrt{||\tilde{x} -  x||_2^2} \label{eq:mse} \\
    \text{SSIM} (\tilde{x}, x) \ &= \ \frac{(2\mu_{\tilde{x}}\mu_x + c_1) (2\sigma_{\tilde{x}}\sigma_x + c_2)}{(\mu_{\tilde{x}}^2 + \mu_x^2 + c_1) ( \sigma_{\tilde{x}}^2 + \sigma_x^2 + c_2)} \label{eq:ssim}
\end{align}

\newpage \cleardoublepage
\section{- Results}
Comparison of all trained models revealed that the networks trained on normalized (by maximum) data performed best. Scaling the data proved to be necessary to have a good generalizing model. Normalization by the maximum absolute value results in scaled data without having altered the physics of the wavefield, something that is no longer true when standardizing the data or normalizing to a custom range. Application of Tukey tapering to the patched data proved to decrease the effect of the undesired edge effects (present in the training data) on the inference results. Therefore, all deep learning results that will follow are based on normalized, tapered models.

\subsection*{Prior space sampling}
As stated before, it is important for a neural network to generalize well to new data. The ability of generalization is determined by the prior space sampled from the training data. The generalization quality of the networks are also dependent on the amount of data used during training because an incorrect ratio between number of training data and number of network parameters could lead to under- or overfitting. First, the effect of data complexity is studied, later the decimation patterns. Varying both of these factors results in a varying amount of training data as well.

Initially, the five different decimation patterns consisted of two irregular and three regular patterns, thereby decimating the data between 50 and 80 \%. Four different models are compared for both the U-Net and RIM, based on different training data consisting of \textit{Gulf (of Suez)} (every second shot), \textit{Rdat} (every second shot of the largest salt dataset), \textit{GulfRdat} (a combination of the former two) or Ocean Turbulence Data (\textit{OTD}; every second shot). The different decimation percentages in addition to patching results in a dataset size of just over 100.000 images for the last two models, just under 100.000 for only \textit{Rdat} and only around 10.000 for \textit{Gulf}. 75 percent of these images went into training, the other 25 percent is used for testing and validation. 

\subsubsection*{Data complexity}
\begin{figure}[htb!]
\centering
\begin{subfigure}[b]{\textwidth}
   \includegraphics[width=1\linewidth]{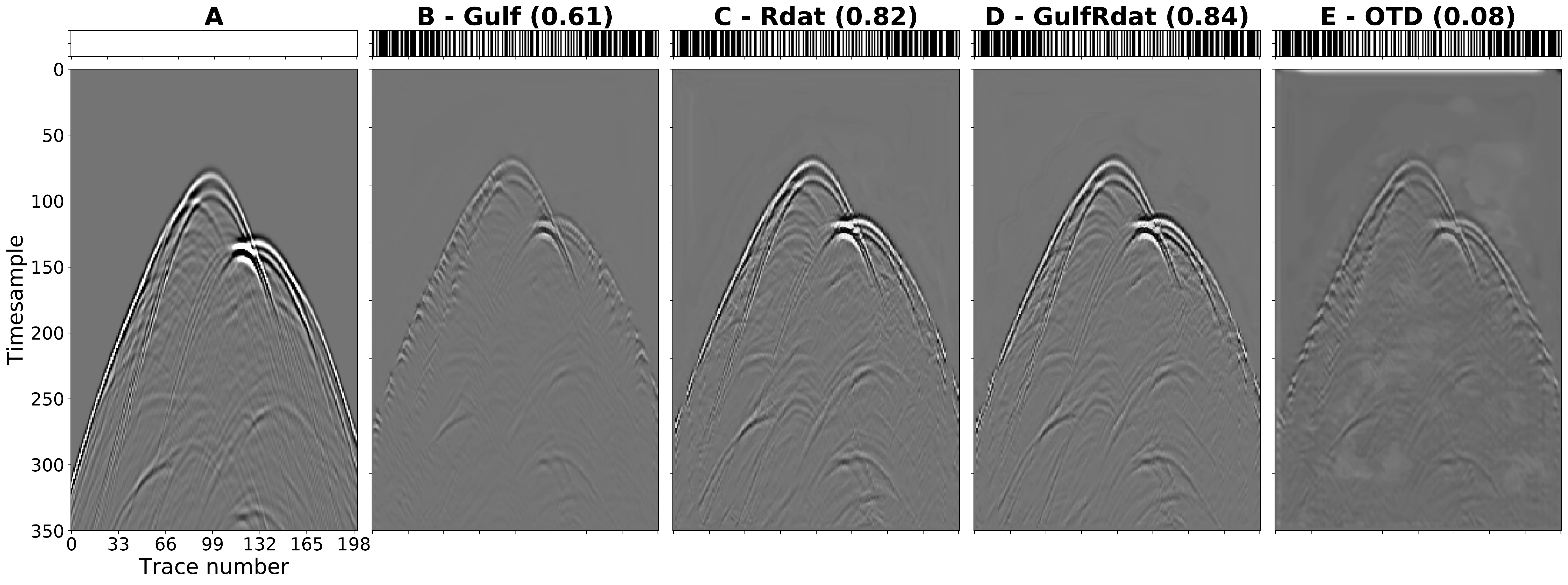}
   \caption{U-Net}
   \label{fig:mc_unet} 
\end{subfigure}

\vspace{\floatsep}

\begin{subfigure}[b]{\textwidth}
   \includegraphics[width=1\linewidth]{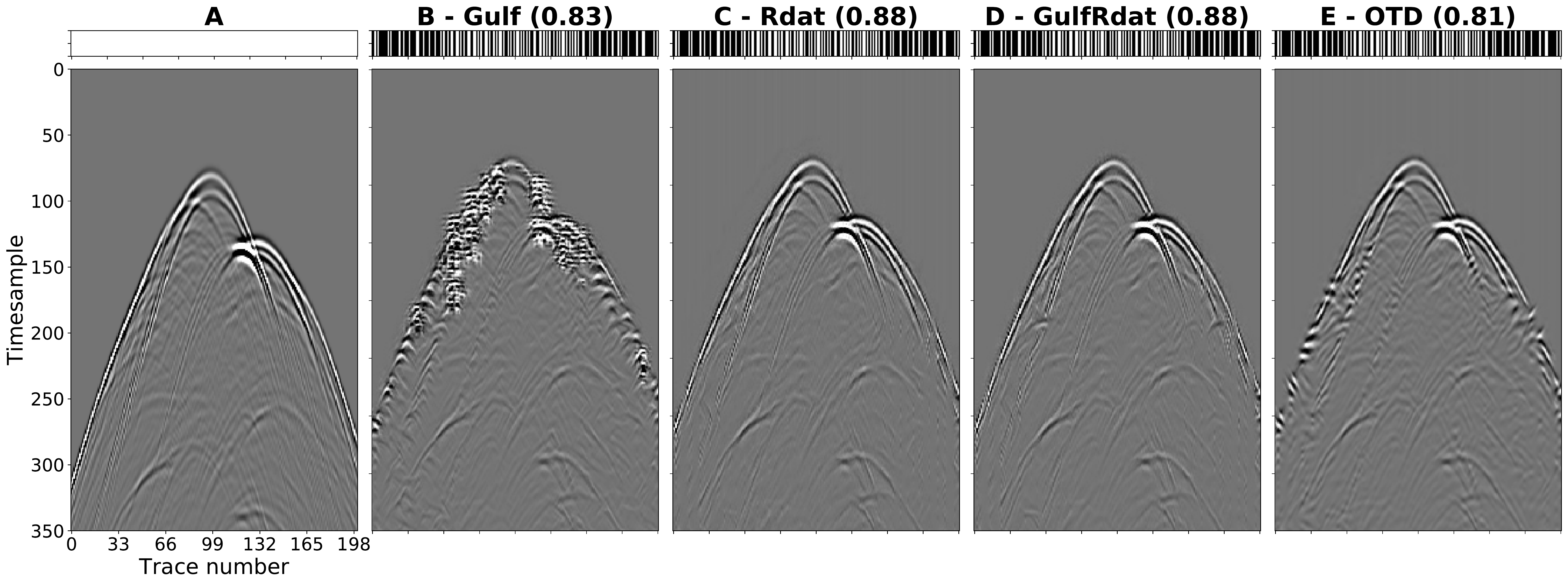}
   \caption{RIM}
   \label{fig:mc_rim}
\end{subfigure}

\caption{Reconstruction of 62 \% irregularly decimated shot gather from a complex single shot salt dataset (normalized). The top bar represents the decimation pattern in which black stands for missing trace. None of the models in panels B-E have been trained on this shot gather. Panel A represents the original shot gather that has to be inferred by the network, therefore the reconstructions in panel B-E should be as close as possible to this panel if inference were perfect. The quality of generalization to unseen data is indicated by the SSIM in brackets and the amplitude of the reconstruction.}
\label{fig:mc}
\end{figure}

Table \ref{tab:mc} in combination with figure \ref{fig:mc} illustrate the effect of data complexity on the potential of the networks to generalize to unseen data. From the average SSIM in table \ref{tab:mc} (arithmetic mean of all but training data performance), it can be deduced that all models perform best on their training data and that the RIM overall performs slightly better than the U-Net. The RIM generalizes equally well with models trained on different higher complexity datasets and poorer when inference is performed on data with a higher complexity than seen during training. This result is to be expected as based on the data complexity discussion given before. U-Net on the other hand, has more trouble generalizing to unseen datasets especially if trained on only the ocean turbulence data that consists of many diffractions and reflections but very little velocity variations (and therefore very little slope variation).

Figure \ref{fig:mc} illustrates this effect and now also gives an indication of the misfit between the network's inference results and to be inferred dense data. The displayed shot gather comes from the single shot salt dataset (\textit{Pdat}) that none of the models had been trained on. This dataset is different from the rest because the data is generated from a source within the medium. The decimation is irregular with a percentage of 62 \% (within the range of decimation percentages in the training data). The 8 different reconstruction panels (B-E in figure \ref{fig:mc_unet} and \ref{fig:mc_rim}) are all very different. For example both reconstructions made by the network trained on \textit{Gulf}-data only, show many small-scale structures on the left flank than present in the dense data (see panels B in figure \ref{fig:mc}). In the RIM it is clear that many small-scale structures, most likely related to the curvature in the training data, overprint the desired curvature of the salt data. In the U-Net this effect is less pronounced, related to the fact that that network also underestimates the amplitude of the reconstruction. Both networks perform best when trained on a combination of complex salt dataset and the Gulf of Suez dataset that includes many velocity and slope variations. 

\begin{table}[h]
\centering
\begin{tabular}[width=\textwidth]{l|llllllll}
 & \multicolumn{2}{c}{\textbf{Gulf}} & \multicolumn{2}{c}{\textbf{Gulf / Rdat}} & \multicolumn{2}{c}{\textbf{Rdat}} & \multicolumn{2}{c}{\textbf{OTD}} \\
& U-Net & RIM & U-Net & RIM& U-Net & RIM& U-Net & RIM \\
\hline
\textbf{Gulf} & \cellcolor[HTML]{C0C0C0}{\color[HTML]{333333} 0.88} & \cellcolor[HTML]{C0C0C0}{\color[HTML]{333333} 0.92} & \cellcolor[HTML]{C0C0C0}{\color[HTML]{333333} 0.89} & \cellcolor[HTML]{C0C0C0}{\color[HTML]{333333} 0.90} & 0.83 & 0.85 & 0.20 & 0.88 \\
\textbf{Rdat} & 0.77 & 0.77 & \cellcolor[HTML]{C0C0C0}{\color[HTML]{333333} 0.84} & \cellcolor[HTML]{C0C0C0}{\color[HTML]{333333} 0.87} & \cellcolor[HTML]{C0C0C0}{\color[HTML]{333333} 0.82} & \cellcolor[HTML]{C0C0C0}{\color[HTML]{333333} 0.86} & 0.11 & 0.80 \\
\textbf{OTD} & 0.64 & 0.78 & 0.75 & 0.83 & 0.75 & 0.81 & \cellcolor[HTML]{C0C0C0}{\color[HTML]{333333} 0.21} & \cellcolor[HTML]{C0C0C0}{\color[HTML]{333333} 0.91} \\
\textbf{Pdat} & 0.63 & 0.65 & 0.75 & 0.79 & 0.73 & 0.77 & 0.13 & 0.74 \\
\hline
\multicolumn{1}{|c}{\textit{\textbf{Average}}} & 0.68 & \textbf{0.73} & 0.75 & \textbf{0.81} & 0.77 & \textbf{0.81} & 0.15 & \multicolumn{1}{c|}{\textbf{0.81}} \\
\hline
\end{tabular}
\caption{Average SSIM for inference using the trained models (columns) on the to be inferred dense data (rows). The SSIM are computed as an arithmetic mean over the SSIM for 10 different decimation percentages (5 regular, 5 irregular) for 3 shot gathers in the data (if available; left quarter, center, right quarter) without taking the training data into the calculation (indicated by gray cells). All models perform best on the data they are trained on and the RIM outperforms the U-Net in these tasks.}
\label{tab:mc}
\end{table}

\subsubsection*{Decimation patterns}
The networks were initially trained on 5 different decimation masks, ranging in decimation percentage between 50 and 80 \%. From these patterns, 2 were irregular and 3 regular. When performing inference on data decimated by between 25 and 82 percent it is observed that the networks can generalize better to lower percentages than towards the higher end of the range present in the prior space. This means that the reconstruction quality thus decreases when the data is highly decimated. There is no clear indication that the networks perform better on irregular or regular decimated data, unlike in the deterministic inversion that tends to be able to reconstruct irregularly sampled data better. 
Training the RIM on only two patterns (50 \% regular and 84 \% irregular) in the same prior space range resulted in similar observations. Using more patterns in the same range (50, 67, 75, 80 \% regular and 75, 81, 2x 84 \% - random jittering- irregular) improved the reconstruction quality but increased the training time as more training data became available. With less decimation patterns in the lower range, the RIM could generalize well to percentages just outside this training range and predicted structures similar to the training data at much higher percentages. These observations thus explain that it is important to train the network in the range of decimation percentages that have to be inferred during training to reconstruct well. The U-Net performance varied highly with a change in training data and that resulted in a performance drop when less data or decimation percentages are used. \\

\noindent Based on the previous discussion on prior-space sampling, the networks trained on half of the Gulf of Suez and salt data for five different decimation percentages (previously called \textit{GulfRdat}) are selected for further inference. This is a weigh-off between training time and inference performance at different percentages. Training the RIM for 40 epochs using just over 100.000 training images on a single GPU took around 12 hours. The U-Net is not a recurrent neural net and requires less memory of the GPU, training this network on the same data and number of epochs took only 1.5 hours. Performing inference on a single full-size shot gather is almost instantaneous, whereas deterministic inversion can take minutes per gather before convergence is reached. 

\subsection*{2D gather reconstruction}
\begin{figure}[htb!]
    \centering
    \includegraphics[width=\textwidth]{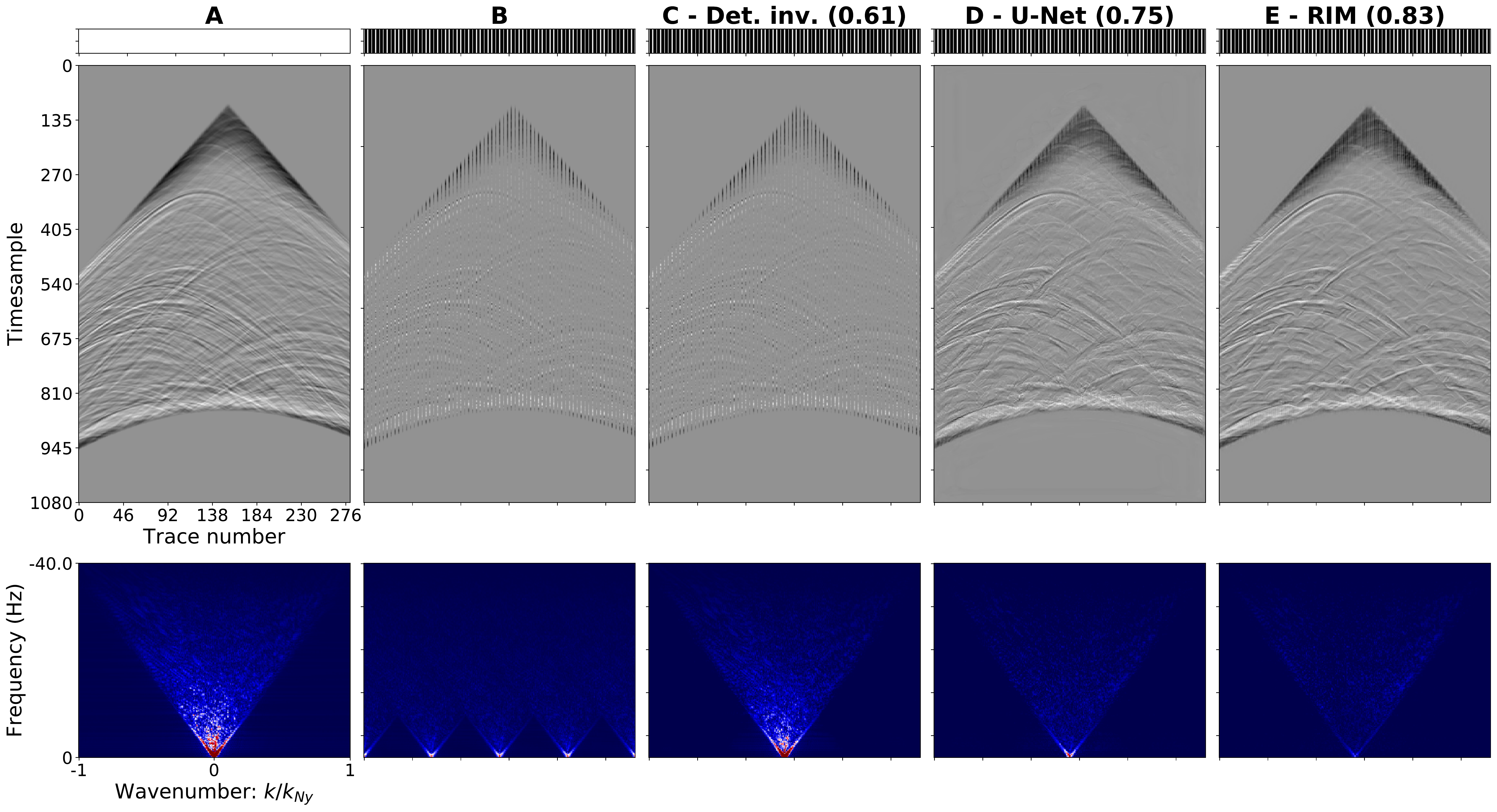}
    \caption{The reconstruction of a central shot gather from the ocean turbulence dataset; each panel consists (from top to bottom) of a bar representing sample distribution (black for missing, white for sampled trace), the normalized wavefield and the corresponding Fourier spectrum. A) Original dense seismic gather, no missing data. B) Data regularly decimated by factor 4 (75 \%), spatial aliasing in the Fourier domain occurs. C - E) Reconstruction using three different approaches, Fourier spectrum is misfit to A. In brackets the SSIM between A and the reconstruction in space-time domain is given. The RIM reconstruction has the lowest misfit in space-time and frequency-wavenumber domain as well as the highest SSIM.}
    \label{fig:otd_2d}
\end{figure}

The reconstruction results for a central shot gather from the ocean turbulence dataset are shown in figure \ref{fig:otd_2d}. Panel A illustrates the temporal bandwidth and spatial variation present in the ocean turbulence dataset. The first arrivals have a strong amplitude, later arrivals are less pronounced but because of normalization and filtering still clearly visible. In this example, the shot gather is regularly decimated by factor 4, resulting in the decimated gather of panel B. Because of sub-nyquist spatial decimation, spatial aliasing in the frequency-wavenumber domain occurs as can be seen in the corresponding Fourier spectrum.

Solving the deterministic inversion without regularization results in panel C of figure \ref{fig:otd_2d}. By visual inspection and comparison of the norms in table \ref{tab:rmse_2d}, there is no difference between the decimated and the reconstructed gather. The misfit between the original Fourier domain image and the Fourier transform of the reconstruction equals the original Fourier domain image. This means that the inversion is not capable of reconstructing the 75 \% missing seismic traces eventhough the iterative inversion has converged. Both deep learning approaches on the other hand, panel D and E in figure \ref{fig:otd_2d}, are capable of reconstructing the missing seismic data. In both panels there is still an imprint of the missing traces, this is especially clear in the first arrivals. The later reflections and diffractions seem to not have this imprint resulting in a low misfit in both the spatial and Fourier domain. Similar as what has been observed before, the U-Net introduces low frequency structures into the reconstruction visible in the low frequency, low wavenumber part of the misfit that has a higher amplitude than that same area for the RIM's reconstruction. The U-Net again also underestimates the amplitude of the data more than the RIM (see the difference in norms in table \ref{tab:rmse_2d}). The training data included higher velocity variations than present in the to be inferred data as well as structural variation. This, structure-wise, results in a high correspondence of the predicted wavefields and the dense wavefield (to be inferred). Not just the strong first arrivals, but also the later diffractions and reflections are reconstructed without loss of bandwidth. 

Both deep learning approaches are thus capable of reconstructing the missing data to similar extent, thereby decreasing spatial aliasing in Fourier domain. The higher SSIM values and lower misfit amplitudes of RIM reconstructions are not limited to this specific gather or dataset only, table \ref{tab:rmse_2d} indicates that this is a general trend. The presented results are based on 75 \% regularly decimated data and can be generalized to other gathers and decimation percentages as well. Where the deterministic inversion on the decimated data and the forward decimation operator already breaks down at very low decimation percentages due to the Shannon-Nyquist sampling theorem, the neural nets performance only start to decrease at decimation percentages near the edge of the sampled prior space. 

\begin{table}[htb!]
\centering
\begin{tabular}[width=\textwidth]{l|cc|cc|cc|cc}
     & \multicolumn{2}{c}{\textbf{Gulf - 27}} & \multicolumn{2}{c}{\textbf{Rdat - 97}} & \multicolumn{2}{c}{\textbf{Pdat - 1}} & \multicolumn{2}{c}{\textbf{OTD - 154}} \\
     & \multicolumn{2}{c}{(4.292e3)} & \multicolumn{2}{c}{(0.938)} & \multicolumn{2}{c}{(29.08e3)} & \multicolumn{2}{c}{(4.242e-3)} \\
     & SSIM & norm & SSIM & norm & SSIM & norm & SSIM & norm \\
     \hline 
     \textbf{Decimated gather} & 0.82 & 2.151e3 & 0.77 & 0.474 & 0.67 & 14.64e3 & 0.61 & 2.131e-3\\
     \textbf{Inversion} & 0.82 & 2.151e3 & 0.77 & 0.474 & 0.67 & 14.64e3 & 0.61 & 2.131e-3 \\
     \textbf{U-Net} & 0.87 & 3.070e3 & 0.86 & 0.706 & 0.79 & 20.59e3 & 0.75 & 3.370e-3 \\
     \textbf{RIM} & 0.88 & 3.528e3 & 0.90 & 0.821 & 0.81 & 25.73e3 & 0.83 & 3.915e-3 
\end{tabular}
\caption{A comparison of the reconstruction for the different approaches to inversion. The different gathers are regularly decimated by a factor 4 (75 \% decimation), the norm of the dense shot gathers is given in brackets after the name of the dataset and the selected shot. The deterministic iterative inversion cannot solve the reconstruction problem for all datasets at this decimation percentage (no difference between input decimated gather and reconstruction), the RIM slightly outperforms the U-Net when comparing the metrics.}
\label{tab:rmse_2d} 
\end{table}

\subsection*{3D Ocean Turbulence reconstruction}
Because the neural nets are trained to reconstruct 2D seismic gathers, a two-step inference procedure is followed to reconstruct the 3D decimated dataset. The total 3D reconstruction is thus an inference result created by first reconstructing all shot gathers and, after sorting to common receiver gathers, in a second step the receiver gathers. The deterministic inversion uses the forward operator and performed a 1000 iterations. Next to that, for the 3D inversion it is assumed that the first-, second-order as well as the spatial cross-derivatives are available therefore taking more data into the inversion and solving a multichannel reconstruction problem. The data are decimated by 94 \% resulting in randomly missing about every fourth trace in both the source and receiver dimension. In total there is 16 times less data present in the decimated wavefield than in the dense wavefield. The decimation pattern is equal in the source and receiver domain, source and receiver positions are colocated in this dataset. Therefore each position will have either both shot and receiver sampled or none of them.

Table \ref{tab:met_3d} compares the inference and inversion results for 5 different methods. Because of the two-step procedure used in inference, the two different networks (RIM and U-Net) can also be used jointly such that the networks could benefit from each others reconstruction made in the first step. The best overall reconstruction is clearly made by the deterministic inversion that used the forward operator, the decimated data and all eight (cross-)derivatives. All deep learning methods however, still estimate the wavefield in a decent matter considering the fact that these networks only know the decimated data and, in case of the RIM, a 2D version of the forward operator. Because two steps are taken in the inference procedure, the second inference step takes place on reconstructed data, this reconstruction is far from perfect and therefore error propagation occurs. From table \ref{tab:met_3d} it should be clear that the reconstruction is best at positions where some data was sampled. Because of the used loss function in training, the networks are free to alter also the traces that where sampled instead of only the missing traces. The inversion uses the forward operator and does not allow the alteration of sampled traces, therefore the misfit between the inference results could always be higher than that of the inversion.  

\begin{table}
\centering
\begin{tabular}[width=\textwidth]{l|cc|cc|cc}
     & \multicolumn{2}{c}{\textbf{Decimated data}} & \multicolumn{2}{c}{\textbf{Sampled data}} & \multicolumn{2}{c}{\textbf{Total 3D dataset}} \\
     & \multicolumn{2}{c}{(44.91)} & \multicolumn{2}{c}{(26.07)} & \multicolumn{2}{c}{(51.94)} \\
     & SSIM & norm & SSIM & norm & SSIM & norm \\
     \hline 
     \textbf{Inversion} & 0.82 & 33.57 & 0.86 & 22.14 & 0.84 & 40.21 \\
     \textbf{RIM} & 0.71 & 20.23 & 0.78 & 17.36 & 0.73 & 26.66 \\
     \textbf{U-Net / RIM} & 0.68 & 17.96 & 0.77 & 15.89 & 0.70 & 23.98 \\
     \textbf{U-Net} & 0.64 & 16.86 & 0.75 & 17.25 & 0.67 & 24.12 \\
     \textbf{RIM / U-Net} & 0.65 & 16.79 & 0.78 & 17.45 & 0.69 & 24.21
     \end{tabular}
\caption{A comparison of 3D inversion results for the 94 \% decimated ocean turbulence data. The deterministic inversion in this case performs best on all components. The two-step RIM reconstruction again estimates the amplitudes of the reconstruction better than the U-Net. Combining the U-Net and RIM leads to a better 3D reconstruction than using the U-Net for two steps, possibly because the RIM uses the forward operator in the estimation. The norm of the original part of the data is given in brackets, all norms are scaled by factor 1e3.}
\label{tab:met_3d}     
\end{table}

\begin{sidewaysfigure}[htb]
   \includegraphics[width=\textwidth]{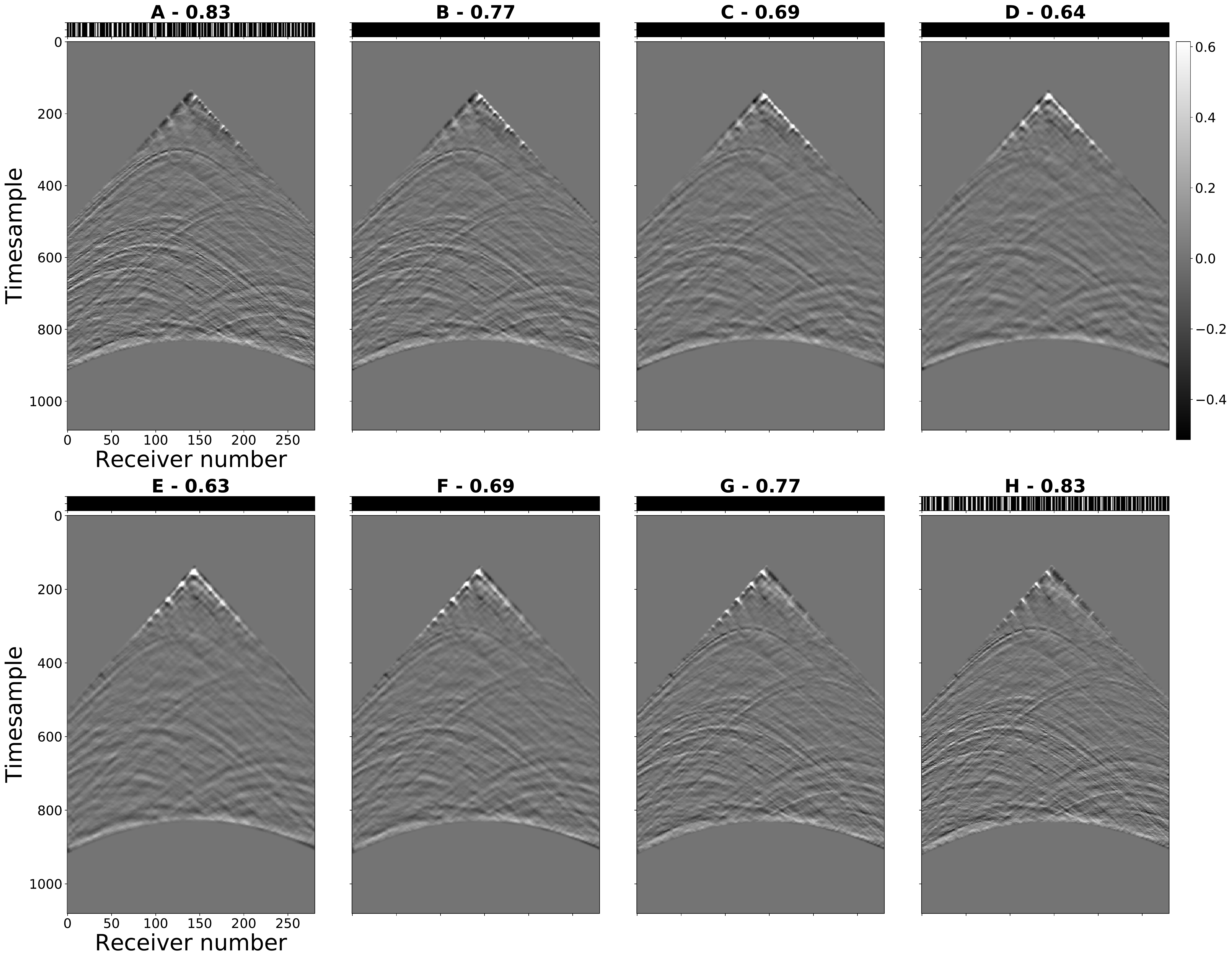}
   \caption{Deterministic inversion results for 94 \% decimated ocean turbulence dataset, shot gather 140 - 147. Bar on top of each panel represents sample distribution (white for sampled, black for decimated), SSIM values are reported for each gather as well. Quality of reconstruction decreases and bandwidth is lost with distance from last sampled gather.}
\label{fig:3dI_shot}
\end{sidewaysfigure}

\begin{sidewaysfigure}[htb]
   \includegraphics[width=\textwidth]{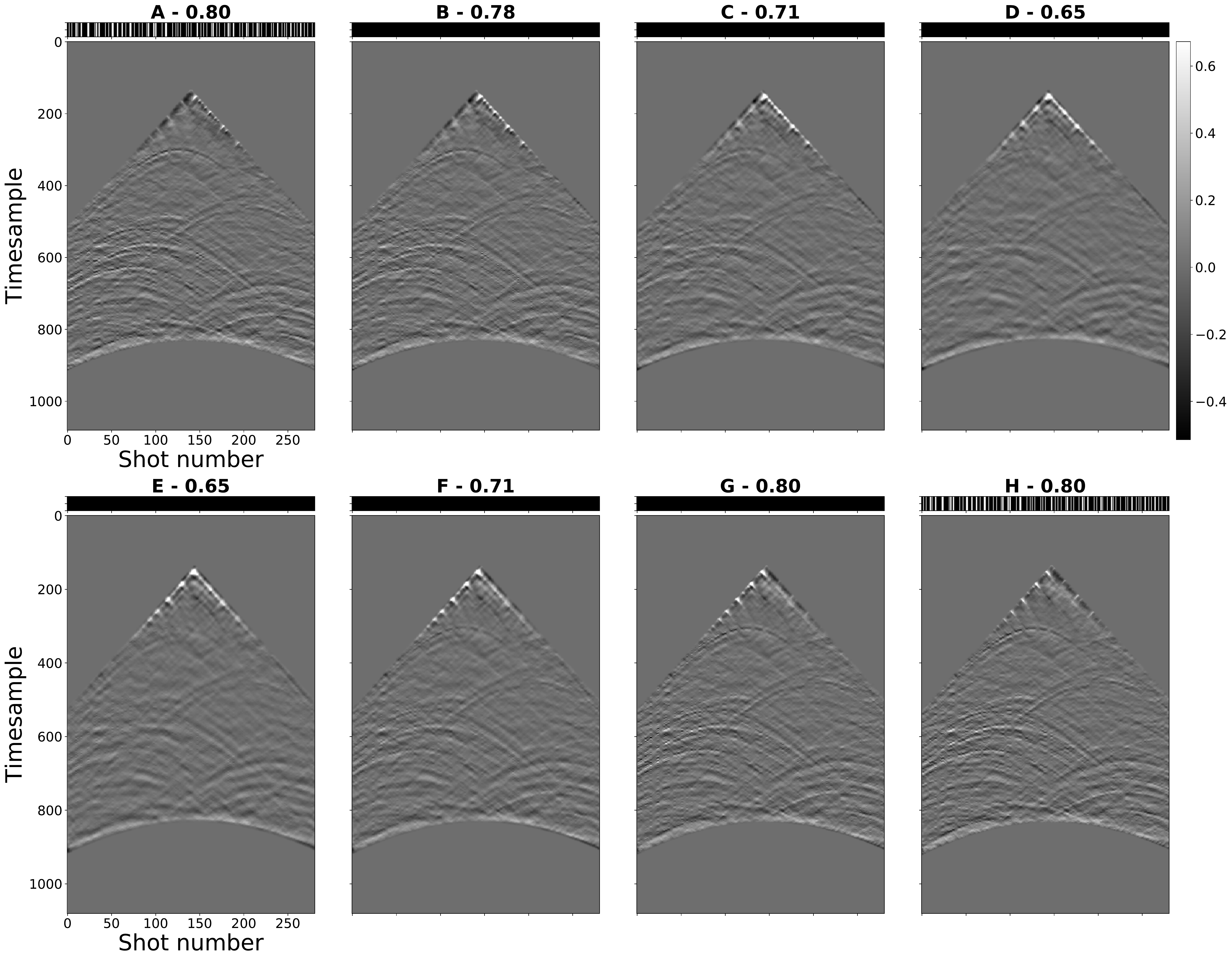}
   \caption{Deterministic inversion results for 94 \% decimated ocean turbulence dataset, receiver gather 140 - 147. Bar on top of each panel represents sample distribution (white for sampled, black for decimated), SSIM values are reported for each gather as well.}
\label{fig:3dI_rec}
\end{sidewaysfigure}

\begin{sidewaysfigure}[htb]
   \includegraphics[width=\textwidth]{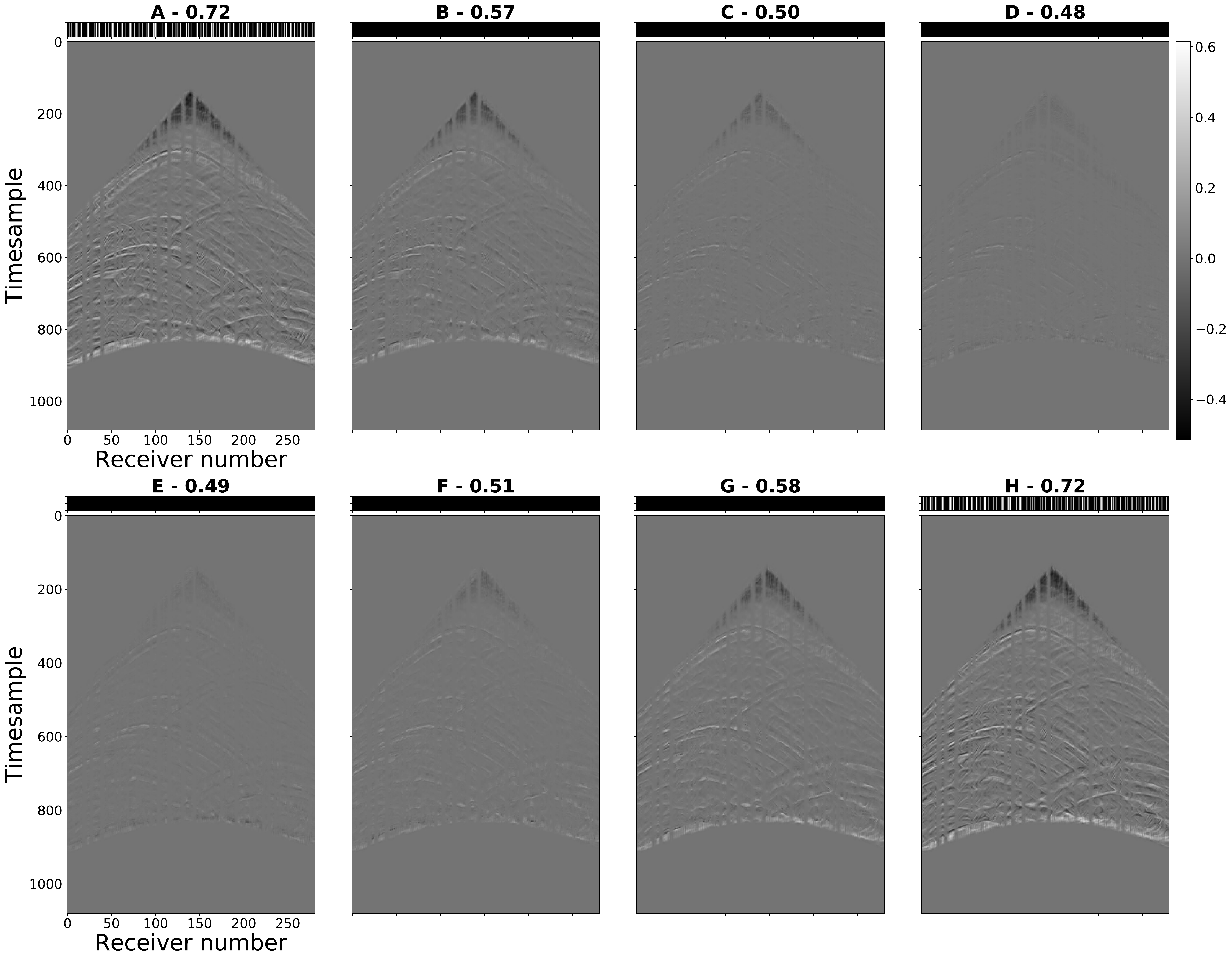}
   \caption{Two-stage RIM inference results for 94 \% decimated ocean turbulence dataset, shot gather 140 - 147. Bar on top of each panel represents sample distribution (white for sampled, black for decimated), SSIM values are reported for each gather as well. Reconstruction is poorer than the deterministic inversion, reconstruction quality decreases with distance from last sampled shot yet there is no clear indication of loss of bandwidth.}
\label{fig:3dR_shot}
\end{sidewaysfigure}

\begin{sidewaysfigure}[htb]
   \includegraphics[width=\textwidth]{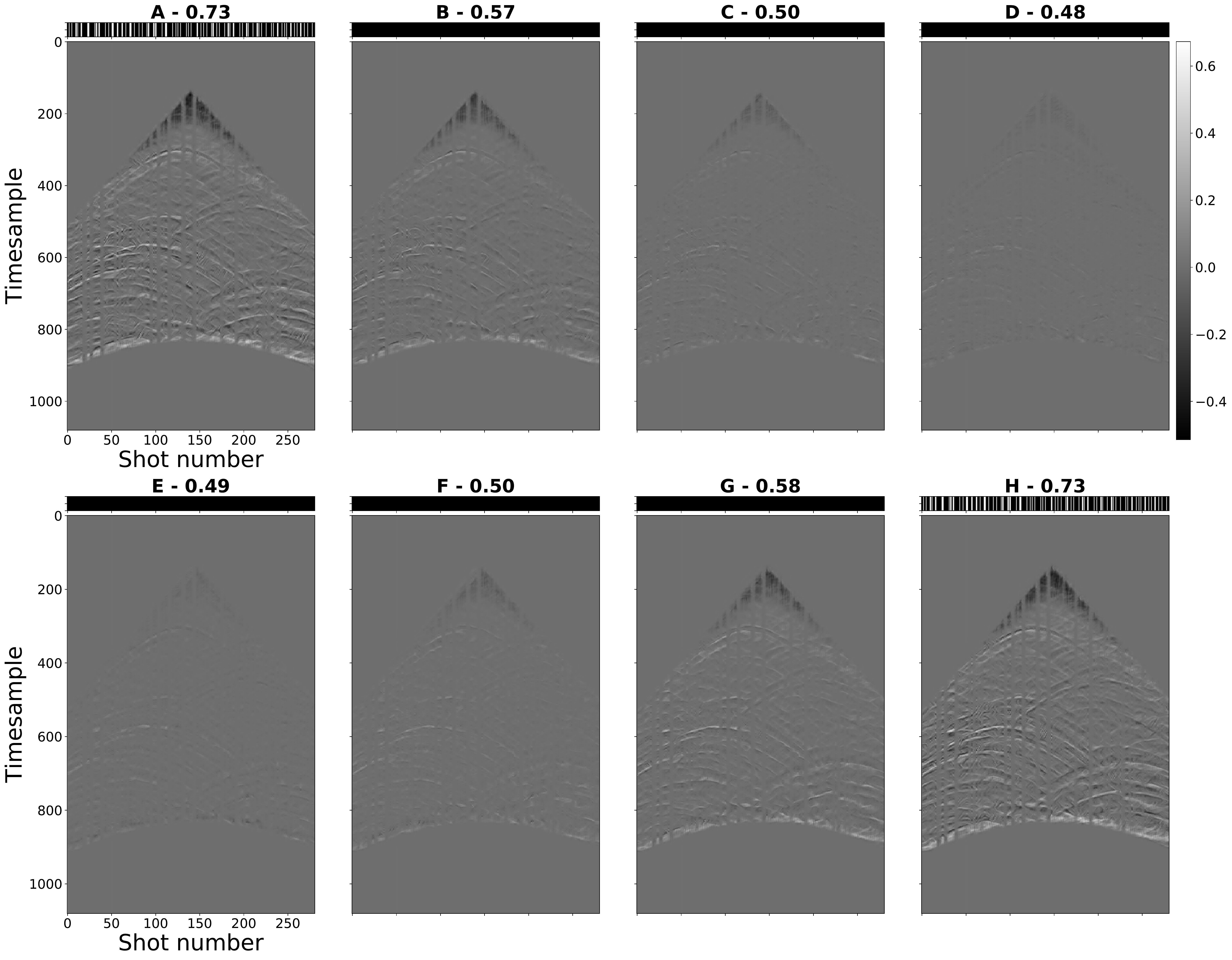}
   \caption{Two-stage RIM inference results for 94 \% decimated ocean turbulence dataset, receiver gather 140 - 147. Bar on top of each panel represents sample distribution (white for sampled, black for decimated), SSIM values are reported for each gather as well.}
\label{fig:3dR_rec}
\end{sidewaysfigure}

Figures \ref{fig:3dI_shot} and \ref{fig:3dI_rec} display the dense wavefield estimates from deterministic inversion for a set of shots in the center of the ocean turbulence dataset. These results are compared to the best probabilistic estimate of the wavefield made by the RIM in figures \ref{fig:3dR_shot} and \ref{fig:3dR_rec}. Because the data is randomly decimated by 75 \% over each dimension, the maximum amount of missing traces in a row within a gather corresponds to six. In the panels of all figures, only the first and last shot/receiver were recorded and the traces in all other missing shots/receivers are reconstructed. Because the RIM reconstructs the decimated data in two steps over the two dimensions, the maximum decimation percentage the network takes as an input equals that of the single dimension, this 75 \% decimation falls just within the range sampled by the prior space.  

Traces in the six missing shots in figure \ref{fig:3dI_shot} are reconstructed by the deterministic inversion. From all approaches, the amplitude of this reconstruction best approximates the dense wavefield. The misfit increases further away from the last sampled shot, yet all major seismic events are accurately recovered. In panel A-D it can be observed that the temporal bandwidth of the reconstruction also decreases with distance from the last sampled shot. As expected, more densely sampled areas result in a better reconstruction. The same general trend can be observed in figure \ref{fig:3dI_rec} for the missing receivers because the decimation patterns over both dimensions are equal and the deterministic inversion method included the 3D forward decimation operator.  

Traces in the six missing shots in figure \ref{fig:3dR_shot} are reconstructed by the two-step RIM inference procedure. Again, the misfit increases further away from the last sampled shot. The temporal bandwidth of the reconstruction however does not seem to decrease with distance, this approach does underestimate the dense wavefield amplitude however. At source and receiver locations where many datapoints are missing, the imprint of the decimation pattern is more evident than in the deterministic inversion. The RIM reconstruction is relatively poor in panels D and E, where the distance to the last sampled shot/receiver is largest. This is most likely due to the fact that as an input to the model, these panels had no data. The reconstruction is thus fully based on inference and the build up of errors over the two steps. 

\newpage \cleardoublepage
\section{- Discussion}
In order to solve the reconstruction problem, two different approaches have been studied. The wavefields reconstructed with the use of deterministic inversion without regularization, verify Shannon-Nyquist sampling theorem that states that dense wavefields can be reconstructed from the decimated (sampled) wavefields only if the sampling frequency is not less than twice the Nyquist frequency. \cite{herrmann2010randomized} studied the effect of different decimation patterns on the imprint in the Fourier spectrum. Regular sampling will lead to sparse and strong aliased signals in the Fourier spectrum where irregular sampling tends to generate weaker decimation artifacts. The regular sampling artifacts hinder the reconstruction and dominate the misfit, whereas the irregular sampling artifacts are less distinct and therefore do not hinder the reconstruction of the original main structures in the wavefield. Because of irregularities or limitations in data acquisition, sampled data are often not fulfilling the sampling criterion and therefore aliasing occurs. These effects are also observed in this study. At lower decimation percentages the deterministic inversion can reconstruct the data for both regular and irregular decimated data. The best reconstructions are made on irregularly decimated data. However, for higher decimation percentages the inversion without regularization is not able to solve the inverse problem for both regular and irregular decimation. Deterministic inversion is only limited to very low decimation percentages, yet it would be beneficial to reconstruct data that is far sparser than reconstructable with the help of inversion. Here, two deep learning approaches have been introduced that have shown to be able to map decimated wavefields into denser wavefields for both regular and irregular, highly sparse data. 

\subsection*{Deterministic versus Probabilistic approach}
Deep learning approached the inverse problem in a probabilistic sense in which the prior has shown to be of crucial importance. The quality of the reconstruction is mainly dependent on the information extracted from the training data. Sampling the training data results in a prior space distribution that is used in the neural networks inference stage. In the seismic reconstruction problem the most important elements the prior space should contain include reflections and diffractions due to spatial variation, bandwidth, slope variations due to velocity variations and a range of decimation percentages. Unlike the deterministic inversion of 2D decimated gathers, that can only reconstruct data accurately when the sampling criterion is fulfilled, the neural networks have proved to be able to reconstruct 2D seismic gathers with decimation percentages up to the edge of the decimation range the networks were trained on. 

When the derivatives of the data are available however, the deterministic inversion of the reconstruction problem turns into the multichannel reconstruction problem. In this case the deterministic inversion improved as more sparse data could be reconstructed. In the 3D highly sparse reconstruction of ocean turbulence data, the deep learning methods have proved to be able to reconstruct the sparse data without the need of derivatives. The reconstruction quality is not as good as the inversion however but it is believed that the reconstruction can be improved by more extensive training on highly sparse data or creating a neural network capable of taking N-dimensional data as in- and output. The two-step inference procedure is prone to error propagation, something that does not occur when having N-dimensional data as input. The loss of bandwidth in the inversion with distance to last sampled shot is not observed in the inference results, indicating that the used training data was sufficient to describe the bandwidth in the ocean turbulence dataset. Because the extra data taking into the inversion (derivatives) is often not available, deep learning should be considered a viable option in data reconstruction. 

Next to the fact that the deep learning methods do not require anything but the data and possibly the forward operator, another advantage of using deep learning methods over deterministic methods lies in the short inference times. Of course, training a neural network takes time. In the case of the used RIM that corresponds to 12 hours where the U-Net did this in under 2 hours. However, with a good generalizing ability, a network only has to be trained once and can be used for inference on unseen datasets afterwards. The reconstruction of a single 2D seismic gather by inference is almost instantaneous whereas the inversion can take up to minutes per gather. When including the derivatives into the inversion this may take even longer (the 3D inversion in a 1000 iterations took over 14 hours to converge). The training time of neural networks could possible be reduced, based on the discussion of prior space sampling required for a good generalizing model. 

The requirement of having a large training data to extract an accurate description of the prior space, could be seen as a difficulty in deep learning as well. In this case, the training data are created synthetically from dense seismic wavefields that include a range of different properties and structures. This means that in all cases it is best to either use existing dense data for training or to sample part of the acquisition densely, thereby providing a possibility of generating synthetic training data consisting of structures present in the to be reconstructed data. As noticeable in the results, without accurate prior space sampling the deep learning networks cannot generalize well enough. Of course, the required quality of the reconstructed data also depends on what this data will be used for in post-processing steps. For example, migration is less demanding than full waveform inversion that attempts to use every single arrival. Therefore making exact conclusions based on the presented metrics here should be done with care, taking the ultimate aim of the reconstruction into account. In seismics, collecting suitable and enough training data should be a manageable task as the required features are very common in seismic data. 

\subsection*{Comparison of deep learning methods}
The two deep learning architectures used here are the Recurrent Inference Machine (RIM) and U-Net. Both methods require training data to update their internal state to match the approximate inverse of the forward operator that generated the decimated data. The RIM approaches inverse problems by combining the known forward operator and the sampled data within a main Recurrent Neural Net (RNN) cell. According to \cite{putzky2017recurrent}, this approach that combines the inference and training stage is crucial and unique to solving inverse problems with deep learning. That the RIM has to potential to solve inverse problems has been demonstrated here by solving the reconstruction problem for which the forward operator is the linear (computationally inexpensive) restriction operator. The RIM demonstrated to generalize well to unseen data and decimation percentages also with a limited amount of training data. From the results it can be concluded that the RIMs have a low tendency to overfit the training data while generalizing well outside the prior range. 

That the RIM is not the only neural net that can represent the inverse of the restriction operator, has been proven with the help of the U-Net. Like the RIM, the U-Net makes use of convolutional operators to extract higher-level features from the input data. However, the U-Net does not use a RNN or the forward operator. In both the 2D seismic gather and the 3D highly decimated reconstruction, the U-Net consistently underestimates the amplitude of the reconstruction and introduces lower frequency structures in the prediction. Most often however, it is possible to filter these lower frequency structures from the predictions and reach results that are similar to the predictions made by the RIM. Likewise, it is often not the absolute amplitude of the reconstruction that is the main goal, the relative distribution of amplitudes is of higher importance as this is a measure of contrast in subsurface properties. This indicates that structure-wise, the reconstruction of the U-Net after filtering could be good enough for further processing as well. Training the U-Net on different training data resulted in highly varying inference results. It can therefore be concluded that the U-Net is much more likely to overfit the training data, possible because of the high number of trainable parameters in the network, and is therefore more prone to prior space variance.  

During the course of this study, another study has been published by \cite{mandelli2019interpolation} in which the U-Net is again used to solve the reconstruction problem as a pre-processing step before using the reconstructed data for migration. There however, as a post-processing step, at the sampled locations the traces are removed from the network's prediction and replaced by the actual sampled traces. \cite{mandelli2019interpolation} find that the U-Net can be used to solve the reconstruction problem. However, their results are based on decimation percentages 10, 30 and 50. Similar observations of poorer generalization to unseen data or decimation patterns are observed.

When taking these considerations into account it can be stated that the reconstructed wavefields in both 2D and 3D made by the RIM are slightly better (in structural similarity, norm as well as dealiasing) than that of the U-Net while both methods perform better than the single channel deterministic inversion at higher decimation percentages. In this decision, emphasis is put on the fact that the RIM generalizes better to unseen data and decimation percentages outside the prior range. When the deterministic inversion does include the derivatives of the data (multichannel reconstruction), the reconstruction improves and becomes better than deep learning methods. Deep learning has proven to be a promising strategy to the single channel reconstruction problem that does not lose bandwidth over the reconstructions and should be considered in N-dimensional problems as well when only the decimated data is acquired.

The choice of hyperparameters in the RIM architecture is based on considerations made by Patrick Putzky and described in \cite{lonning2018recurrent}. The U-Net architecture is created such that it extracts a similar number of features in the first layer as does the RIM (here 64). The number of pooling layers is chosen to be four such the representation of the input data has a minimum size in the bottleneck layer. The size of the input data (32 receiver samples, 64 timesamples) is based on the memory load on a single GPU. For the RIM, which has a higher memory load than the U-Net, this input size in batches of 32 was the maximum load the single GPU could use. As observed in the results, with this window size the temporal and spatial structures can be captured such that generalization to full (not windowed; inference stage) seismic gathers is possible. To benchmark the U-Net and RIM, the input size in the U-Net is chosen to be equal to that of the RIM eventhough the computational load is much lower for this network and a larger window could have been chosen. The training data is windowed using non-overlapping patches, results in \cite{mandelli2019interpolation} describe that overlapping patches increase the computational load while resulting in only a very limited increase in inference performance. Even though the neural networks have been trained to reach their, as equal as possible, minimum states, the networks should still be compared with care as their architectures are different.

\subsubsection*{Effect of forward operator}
That the RIM takes the forward operator into account is what is believed to make the RIMs approach to inverse problems better than the U-Net. Unfortunately, because that is not the only difference between the two architectures (1. the RIM is a RNN, 2. the RIM is a RNN that uses the forward operator in its update function), it can only be stated with care that the fact the forward operator is used to solve the inverse problem in the RIM is what makes the RIM a better probabilistic inverse problem solver than the U-Net. To exclude the fact that the RNN is what makes the RIM perform better than the U-Net, a neural network is trained using a unit forward operator. In that case, the prediction made by the RIM are worse than that of U-Net. This observation supports the hypothesis and indicates that the differences between the RIM and U-Net indeed come from the fact that the RIM can extract information from the gradient of the log-likelihood for which the forward operator is required. 

\subsubsection*{More complex forward operator}
Eventhough the U-Net performs slightly worse than the RIM, the U-Net is able to represent the inverse to the linear forward operator decimating the data. Because the RIM is mostly designed in an approach to inverse problems, it was expected to outperform the U-Net. The RIM does perform better than the U-Net, but it did not excel in the reconstruction problem. It is believed that the RIM will excel for more complex (possibly even non-linear) forward operators. As a first test closely related to the reconstruction problem, the reconstruction problem was transformed to the Fourier domain. Reconstructing data in space-time domain can be seen as dealiasing the Fourier spectrum that is aliased due to sub-Nyquist spatial sampling. Because of the current limitations by the single GPU setup it was not possible to study this approach to more complex forward operators. This is related to the fact that taking the Fourier transform of a patch of data results in a local Fourier representation of the data instead of the full global spectrum. Training the networks to dealias the local spectrum did not correspond to dealiasing the global spectrum for all given methods and therefore this should be part of future studies. 

\cite{lonning2019recurrent} did use the RIM as an approximate inverse of a more complex forward operator and also compared this to the U-Net. In this case, the data is sampled in image space with decimation taking place in another data space related to the image space by the Fourier transform. Results from \cite{lonning2019recurrent} indicate that indeed it is the RIMs architecture that makes the network a potential inverse problem solver. The RIM generalized better to unseen data, required less training data (less parameters to train) and did not suffer from structural artifacts as generated by the U-Net. Again the U-Net generalized poorly to unseen data or decimation ranges, linked to the number of trainable parameters. 

\subsection*{Limitations \& Future work}
Unlike the deterministic inversion, the networks were free to alter the sampled traces. This might not have been the best approach and should be changed in the future. A weighting factor and the forward operator could be included in the loss function that then emphasizes that the network should reconstruct the decimated traces only. It is believed that this will positively affect the reconstruction results. 

From these results and those in \cite{mandelli2019interpolation}, it became clear that not just the RIM but also the U-Net has the ability to represent the inverse to the restriction operator. Despite currently being limited by the single GPU setup, it would be interesting to test the ability of both networks to represent more complex (possibly non-linear) operators. Results from \cite{lonning2019recurrent} indicate that in that case the RIM will outperform the U-Net. This statement could be studied in the Fourier domain as a follow-up to this study where reconstruction took place in the space-time domain. With the use of multiple GPUs it would be possible to distribute the training data over multiple GPUs without being limited to the window size of 32x64 currently used. This would mean the networks can be trained to dealias the global Fourier spectrum, thereby reducing spatial aliasing and thus reconstructing decimated data in space-time domain. This study, as well comparisons made by e.g. \cite{kim2018geophysical} and \cite{russell2019machine}, indicate that indeed deep learning should be considered as a viable option to solving inverse problems and especially those for which deterministic inversion is not possible.

It would be interesting to use the reconstructed data volumes in post-processing steps. For example, migration can be performed on the 3D reconstructed highly sparse ocean turbulence data volume. At this point, the comparison between the deterministic and probabilistic approach is limited to the reconstructions and after migration it would be possible to see if the methods result in a similar image of the studied subsurface. Therefore a decisive conclusion should not purely be based on the metrics used in this study, different types of effects can or cannot have an effect in post-processing steps and therefore it is difficult to state exactly what makes a reconstructed image 'good'. Using the reconstructed data volumes for migration is currently part of ongoing studies. 

\newpage \cleardoublepage
\section{- Conclusions}
In this study two different approaches to solving the reconstruction problem, as an example of an inverse problem for which the forward operator is known, have been studied. The deterministic inversion without regularization is not capable of reconstructing the decimated seismic data when the acquisition did not follow the setup specified by Shannon-Nyquist sampling theorem. 

Two deep learning methods, that approach the inverse problem in a probabilistic sense, have been compared on different reconstruction tasks. It can be concluded that the most important element in building a well generalizing neural network is the prior space. In the seismic data reconstruction problem, this prior space should consist of similar features as those to be inferred including bandwidth, structural and velocity variations, and a range of decimation percentages. The ability of the deep learning methods to represent the inverse of the restriction operator is better than that of the deterministic inversion. The predictions made by the network result in higher SSIM values and better estimates of the norm for all studied decimation percentages, patterns and datasets. The deep learning methods are capable of eliminating spatial aliasing in the Fourier domain where the inversion cannot undo the aliasing caused by sub-Nyquist spatial sampling. Both deep learning methods have proved to be able to map decimated data into dense seismic data thereby solving the reconstruction problem. The deterministic inversion can be improved by incorporating spatial derivatives. The two-step multichannel reconstruction made by deep learning proved that deep learning should be considered as a viable option for highly sparse, N-dimensional data reconstruction when only the decimated data are acquired. 

The RIM architecture is specifically designed to approximate the inverse of the forward operator and is compared to the U-Net (initially designed for image segmentation). Benchmarking the RIM against the U-Net leads to the conclusion that the RIM generalizes better to unseen decimation percentages and data due to the nature of the architecture in which the reconstruction is regularized by the forward operator. The RIM contains less trainable parameters thereby being less prone to overfitting. For simple linear operators, the U-Net is also capable of inverting the system except underestimating amplitudes and introducing low frequency artifacts thereby requiring further processing before using the data volumes in e.g. migration and full waveform inversion. 

Benchmarking the RIM against other deep learning architectures for more complex forward operators should be the subject of future studies. However, initial results as presented here show that RIMs have great potential in seismic processing problems where determining a complex inverse map to a known forward problem is the goal of inference by machine learning. 

\newpage\cleardoublepage
\bibliographystyle{seg}
\bibliography{literature}

\begin{thebibliography}{}
\itemsep0pt

\bibitem[Dai and Heckel, 2019]{dai2019channel}
Dai, Z., and R. Heckel,  2019, Channel normalization in {C}onvolutional
  {N}eural {N}etwork avoids {V}anishing {G}radients: arXiv preprint
  arXiv:1907.09539.

\bibitem[Goodfellow et~al., 2016]{Goodfellow-et-al-2016}
Goodfellow, I., Y. Bengio, and A. Courville,  2016, {D}eep {L}earning: MIT
  Press.
\newblock (http://www.deeplearningbook.org).

\bibitem[Hennenfent and Herrmann, 2008]{hennenfent2008simply}
Hennenfent, G., and F.~J. Herrmann,  2008, Simply denoise: {W}avefield
  reconstruction via jittered undersampling: Geophysics, {\bf 73}, V19--V28.

\bibitem[Herrmann, 2010]{herrmann2010randomized}
Herrmann, F.~J.,  2010, Randomized sampling and sparsity: {G}etting more
  information from fewer samples: Geophysics, {\bf 75}, WB173--WB187.

\bibitem[Ioffe and Szegedy, 2015]{ioffe2015batch}
Ioffe, S., and C. Szegedy,  2015, Batch normalization: {A}ccelerating deep
  network training by reducing internal covariate shift: arXiv preprint
  arXiv:1502.03167.

\bibitem[Kim and Nakata, 2018]{kim2018geophysical}
Kim, Y., and N. Nakata,  2018, Geophysical inversion versus machine learning in
  inverse problems: The Leading Edge, {\bf 37}, 894--901.

\bibitem[Kingma and Ba, 2014]{kingma2014adam}
Kingma, D.~P., and J. Ba,  2014, Adam: {A} method for stochastic optimization:
  arXiv preprint arXiv:1412.6980.

\bibitem[L{\o}nning et~al., 2018]{lonning2018recurrent}
L{\o}nning, K., P. Putzky, M.~W. Caan, and M. Welling,  2018, Recurrent
  inference machines for accelerated {MRI} reconstruction: Presented at the
  International Conference on Medical Imaging with Deep Learning (MIDL 2018).

\bibitem[L{\o}nning et~al., 2019]{lonning2019recurrent}
L{\o}nning, K., P. Putzky, J.-J. Sonke, L. Reneman, M.~W. Caan, and M. Welling,
   2019, Recurrent inference machines for reconstructing heterogeneous {MRI}
  data: Medical image analysis, {\bf 53}, 64--78.

\bibitem[Mandelli et~al., 2019]{mandelli2019interpolation}
Mandelli, S., V. Lipari, P. Bestagini, and S. Tubaro,  2019, Interpolation and
  denoising of seismic data using convolutional neural networks: arXiv preprint
  arXiv:1901.07927.

\bibitem[Martinez, 2016]{martinez2016overview}
Martinez, M.~T.,  2016, An overview of {G}oogle's {M}achine {I}ntelligence
  {S}oftware {T}ensor{F}low.: Technical report, Sandia National Lab.(SNL-NM),
  Albuquerque, NM (United States).

\bibitem[Ndajah et~al., 2010]{ndajah2010ssim}
Ndajah, P., H. Kikuchi, M. Yukawa, H. Watanabe, and S. Muramatsu,  2010, {SSIM}
  image quality metric for denoised images: Proc. 3rd WSEAS Int. Conf. on
  Visualization, Imaging and Simulation, 53--58.

\bibitem[Paszke et~al., 2019]{paszke2019pytorch}
Paszke, A., S. Gross, F. Massa, A. Lerer, J. Bradbury, G. Chanan, T. Killeen,
  Z. Lin, N. Gimelshein, L. Antiga, et~al.,  2019, Py{T}orch: {A}n imperative
  style, high-performance deep learning library: Advances in Neural Information
  Processing Systems, 8024--8035.

\bibitem[Peng and Vasconcelos, 2019]{peng2019study}
Peng, H., and I. Vasconcelos,  2019, A study of acquisition-related
  sub-sampling and aperture effects on {M}archenko focusing and redatuming,
  {\it in} SEG Technical Program Expanded Abstracts 2019: Society of
  Exploration Geophysicists,  248--252.

\bibitem[Putzky and Welling, 2017]{putzky2017recurrent}
Putzky, P., and M. Welling,  2017, Recurrent inference machines for solving
  inverse problems: arXiv preprint arXiv:1706.04008.

\bibitem[Ravasi and Vasconcelos, 2020]{ravasi2020pylops}
Ravasi, M., and I. Vasconcelos,  2020, Py{L}ops—{A} linear-operator {P}ython
  library for scalable algebra and optimization: SoftwareX, {\bf 11}, 100361.

\bibitem[Ronneberger et~al., 2015]{ronneberger2015u}
Ronneberger, O., P. Fischer, and T. Brox,  2015, U-net: Convolutional networks
  for biomedical image segmentation: International Conference on Medical image
  computing and computer-assisted intervention, Springer, 234--241.

\bibitem[Ruan, 2019]{JMthesis}
Ruan, J.,  2019, Compressive {A}cquisition and {O}cean {T}urbulence {W}avefield
  {R}econstruction: Master's thesis, Utrecht {U}niversity.

\bibitem[Ruan and Vasconcelos, 2019]{ruan2019data}
Ruan, J., and I. Vasconcelos,  2019, Data-and prior-driven sampling and
  wavefield reconstruction for sparse, irregularly-sampled, higher-order
  gradient data, {\it in} SEG Technical Program Expanded Abstracts 2019:
  Society of Exploration Geophysicists,  4515--4519.

\bibitem[Russell, 2019]{russell2019machine}
Russell, B.,  2019, Machine learning and geophysical inversion—{A} numerical
  study: The Leading Edge, {\bf 38}, 512--519.

\bibitem[Siahkoohi et~al., 2018]{siahkoohi2018seismic}
Siahkoohi, A., R. Kumar, and F. Herrmann,  2018, Seismic data reconstruction
  with generative adversarial networks: Presented at the 80th EAGE Conference
  and Exhibition 2018.

\bibitem[Udacity, 2019]{udacitycourse}
Udacity,  2019, Intro to {D}eep {L}earning with {P}y{T}orch,
  \verb+https://www.udacity.com/course/deep-learning-pytorch--ud188+, accessed
  {S}ummer 2019.

\bibitem[Zbontar et~al., 2018]{zbontar2018fastmri}
Zbontar, J., F. Knoll, A. Sriram, M.~J. Muckley, M. Bruno, A. Defazio, M.
  Parente, K.~J. Geras, J. Katsnelson, H. Chandarana, et~al.,  2018, fast{MRI}:
  An open dataset and benchmarks for accelerated {MRI}: arXiv preprint
  arXiv:1811.08839.

\end{thebibliography}
\end{document}